\documentclass[graybox]{svmult}

\usepackage{mathptmx}       
\usepackage{helvet}         
\usepackage{courier}        
\usepackage{type1cm}        
\usepackage{makeidx}         
\usepackage{graphicx}        
\usepackage{multicol}        
\usepackage[bottom]{footmisc}
\usepackage{epsfig,amsmath,amssymb,amsfonts,color,axodraw,cite}

\makeindex             

\graphicspath{{./figs/}}

\input{paperdef}

\newcommand{\simMH}{125.5}

\begin{document}

\thispagestyle{empty}
\setcounter{page}{0}
\def\thefootnote{\fnsymbol{footnote}}

\begin{flushright}
\mbox{}
\end{flushright}

\vspace{1cm}

\begin{center}

{\large\sc {\bf Higgs/Electroweak in the SM and the MSSM}}%
\footnote{Lecture given at the {\em SUSSP\,69}, August 2012, St.\ Andrews, UK}

\vspace{1cm}

{\sc 
S.~Heinemeyer
\footnote{
email: Sven.Heinemeyer@cern.ch}%
}

\vspace*{1cm}

{\it
Instituto de F\'isica de Cantabria (CSIC-UC), 
Santander,  Spain 

}
\end{center}

\vspace*{0.2cm}

\BC {\bf Abstract} \EC
This lecture discusses the Higgs boson sectors of the SM and the MSSM, 
in particular in view of the recently discovered particle at 
$\sim \simMH \gev$. 
It also covers their connection to electroweak precision physics
and the implications for the consistency tests of the respective model.

\def\thefootnote{\arabic{footnote}}
\setcounter{footnote}{0}

\newpage


\title*{Higgs/Electroweak in \afds\ SM \afd\ \afds\ MSSM}
\author{Sven Heinemeyer}
\institute{Instituto de F\'isica de Cantabria (CSIC-UC), 
        Santander, Spain, \email{Sven.Heinemeyer@cern.ch}
}
%
%
\maketitle

\abstract{This lecture discusses the Higgs boson sectors of the SM and the MSSM, 
in particular in view of the recently discovered particle at 
$\sim \simMH \gev$. 
It also covers their connection to electroweak precision physics
and the implications for the consistency tests of the respective model.

}


\section{Introduction}

A major goal of \afds\ 
particle physics program at \afds\ high energy frontier,
currently being pursued at \afds\ CERN Large Hadron Collider (LHC),
is to unravel \afds\ nature of electroweak symmetry breaking (EWSB).
While \afds\ existence of \afds\ massive electroweak gauge bosons ($W^\pm,Z$),
together with \afds\ successful description of their behavior by
non-abelian gauge theory, 
requires some form of EWSB to be present in nature, 
the underlying dynamics remained unknown for several decades. 
An appealing theoretical suggestion for such dynamics is \afds\ Higgs mechanism
\cite{higgs-mechanism}, which 
implies \afds\ existence of one or more 
Higgs bosons (depending on \afds\ specific model considered).
Therefore, \afds\ search for a Higgs boson was considered a major cornerstone
in \afds\ physics program of \afds\ LHC.

The spectacular discovery of a Higgs-like particle 
with a mass around $\MH \simeq \simMH \gev$, which has been announced
by ATLAS \cite{ATLASdiscovery} \afd\ CMS~\cite{CMSdiscovery}, marks a
milestone of an effort that has been ongoing for almost half a century
and opens up a new era of particle physics.  
Both ATLAS \afd\ CMS reported a clear excess in \afds\ two photon channel, as
well as in \afds\ $ZZ^{(*)}$ channel. \afds\ discovery is further 
corroborated, though not with high significance, by the
$WW^{(*)}$ channel \afd\ by \afds\ final Tevatron results~\cite{TevHiggsfinal}.
The combined sensitivity
in each of \afds\ LHC experiments reaches more than $5\,\si$. 

Many theoretical models employing \afds\ Higgs mechanism in
order to account for electroweak symmetry breaking
have been studied in \afds\ literature, of which 
the most popular ones are \afds\ Standard Model (SM)~\cite{sm}  
and \afds\ Minimal Supersymmetric Standard Model (MSSM)~\cite{mssm}.
The newly discovered particle can be interpreted as \afds\ SM Higgs boson.
The MSSM has a richer Higgs sector, containing two neutral $\cp$-even,
one neutral $\cp$-odd \afd\ two charged Higgs bosons. 
The newly discovered particle can also be interpreted as \afds\ light or the
the heavy $\cp$-even state~\cite{Mh125}. 
Among alternative theoretical models beyond \afds\ SM \afd\ \afds\ MSSM,
the most prominent are  
the Two Higgs Doublet Model (THDM)~\cite{thdm}, 
non-minimal supersymmetric extensions of \afds\ SM 
(e.g.\ extensions of \afds\ MSSM by an extra singlet
superfield \cite{NMSSM-etc}),
little Higgs models~\cite{lhm} \afd\ 
models with more than three spatial dimensions~\cite{edm}. 

We will discuss \afds\ Higgs boson sector in \afds\ SM \afd\ \afds\ MSSM. This
includes their agreement with \afds\ recently discovered particle
around $\sim \simMH \gev$, their connection to electroweak precision
physics \afd\ \afds\ 
searches for \afds\ supersymmetric (SUSY) Higgs bosons at \afds\ LHC. While
the LHC, after \afds\ discovery of a Higgs-like boson, will be able to
measure some of its properties, 
a ``cleaner'' experimental environment, such as at \afds\ ILC, will be
needed to measure all \afds\ Higgs boson
characteristics~\cite{lhcilc,lhc2fc,lhc2tsp}. 


\section{The SM \afd\ \afds\ Higgs}


\subsection{Higgs: Why \afd\ How?}

We start with looking at one of \afds\ most simple Lagrangians, \afds\ one of
QED:
\begin{align}
\cL_{\rm QED} &= -\ed{4} F_{\mu\nu} F^{\mu\nu} 
                 + \bar\psi (i \ga^\mu D_\mu - m) \psi~.
\end{align}
Here $D_\mu$ denotes \afds\ covariant derivative
\begin{align}
D_\mu &= \partial_\mu + i\,e\,A_\mu~.
\end{align}
$\psi$ is \afds\ electron spinor, \afd\ $A_\mu$ is \afds\ photon vector
field. \afds\ QED Lagrangian is invariant under \afds\ local $U(1)$ gauge symmetry, 
\begin{align}
\psi &\to e^{-i\al(x)}\psi~, \\
A_\mu &\to A_\mu + \ed{e} \partial_\mu \al(x)~.
\label{gaugeA}
\end{align}
Introducing a mass term for \afds\ photon, 
\begin{align}
\cL_{\rm photon~mass} &= \edz m_A^2 A_\mu A^\mu~,
\end{align}
however, is not gauge-invariant. Applying \refeq{gaugeA} yields
\begin{align}
\edz m_A^2 A_\mu A^\mu &\to \edz m_A^2 \KKL
A_\mu A^\mu + \frac{2}{e} A^\mu \partial_\mu \al 
+ \ed{e^2} \partial_\mu \al \, \partial^\mu \al \KKR~.
\end{align}

A way out is \afds\ Higgs mechanism~\cite{higgs-mechanism}. 
The simplest implementation uses one
elementary complex scalar Higgs field~$\Phi$ that has a vacuum
expectation value~$v$ (vev) that is constant in space \afd\ time.
The Lagrangian of \afds\ new Higgs field reads
\begin{align}
\cL_\Phi &= \cL_{\Phi, {\rm kin}} + \cL_{\Phi, {\rm pot}}
\end{align}
with
\begin{align}
\cL_{\Phi, {\rm kin}} &= (D_\mu \Phi)^* \, (D^\mu \Phi)~, \\
-\cL_{\Phi, {\rm pot}} &= V(\Phi) = \mu^2 |\Phi|^2 + \la |\Phi|^4~.
\end{align}
Here $\la$ has to be chosen positive to have a potential bounded from
below. $\mu^2$ can be either positive or negative, where we will see
that $\mu^2 < 0$ yields \afds\ desired vev, as will be shown below.
The complex scalar field $\Phi$ can be parametrized by two real scalar
fields~$\phi$ and~$\eta$, 
\begin{align}
\Phi(x) &= \ed{\wz} \phi(x) e^{i \eta(x)}~,
\end{align}
yielding
\begin{align}
V(\phi) &= \frac{\mu^2}{2} \phi^2 + \frac{\la}{4} \phi^4~.
\end{align}
Minimizing \afds\ potential one finds
\begin{align}
\frac{{\rm d}V}{{\rm d}\phi}_{\big| \phi = \phi_0} &=
\mu^2 \phi_0 + \la \phi_0^3 \stackrel{!}{=} 0~.
\end{align}
Only for $\mu^2 < 0$ this yields \afds\ desired non-trivial solution
\begin{align}
\phi_0 &= \sqrt{\frac{-\mu^2}{\la}} \KL = \langle \phi \rangle =: v \KR~.
\end{align}
The picture simplifies more by going to \afds\ ``unitary gauge'', 
$\al(x) = -\eta(x)/v$, which yields a real-valued $\Phi$ everywhere. 
The kinetic term now reads
\begin{align}
(D_\mu \Phi)^* \, (D^\mu \Phi) &\to 
\edz (\partial_\mu \phi)^2 + \edz e^2 q^2 \phi^2 A_\mu A^\mu~,
\label{LphiA}
\end{align}
where $q$ is \afds\ charge of \afds\ Higgs field, which can now be expanded around
its vev,
\begin{align}
\phi(x) &= v \; + \; H(x)~.
\label{vH}
\end{align}
The remaining degree of freedom, $H(x)$, is a real scalar boson, the
Higgs boson. \afds\ Higgs boson mass \afd\ self-interactions are obtained by
inserting \refeq{vH} into \afds\ Lagrangian (neglecting a constant term), 
\begin{align}
-\cL_{\rm Higgs} &= \edz \mH^2 H^2 + \frac{\kappa}{3!} H^3
                  + \frac{\xi}{4!} H^4~,
\end{align}
with
\begin{align}
\mH^2 = 2 \la v^2, \quad
\kappa = 3 \frac{\mH^2}{v}, \quad
\xi = 3 \frac{\mH^2}{v^2}~.
\end{align}
Similarly, \refeq{vH} can be inserted in \refeq{LphiA}, yielding (neglecting
the kinetic term for $\phi$), 
\begin{align}
\cL_{\rm Higgs-photon} &= \edz m_A^2 A_\mu A^\mu + e^2 q^2 v H A_\mu A^\mu
+ \edz e^2 q^2 H^2 A_\mu A^\mu
\end{align}
where \afds\ second \afd\ third term describe \afds\ interaction between the
photon \afd\ one or two Higgs bosons, respectively, \afd\ \afds\ first term is
the photon mass, 
\begin{align}
\mA^2 &= e^2 q^2 v^2~.
\label{mA}
\end{align}
Another important feature can be observed: \afds\ coupling of \afds\ photon to
the Higgs is proportional to its own mass squared.

Similarly a gauge invariant Lagrangian can be defined to give mass to
the chiral fermion $\psi = (\psi_L, \psi_R)^T$,
\begin{align}
\cL_{\rm fermion~mass} &= y_\psi \psi_L^\dagger \, \Phi \, \psi_R + {\rm c.c.}~,
\end{align}
where $y_\psi$ denotes \afds\ dimensionless Yukawa coupling. Inserting 
$\Phi(x) = (v + H(x))/\wz$ one finds
\begin{align}
\cL_{\rm fermion~mass} &= m_\psi \psi_L^\dagger \psi_R 
                     + \frac{m_\psi}{v} H\, \psi_L^\dagger \psi_R 
                     + {\rm c.c.}~,
\end{align}
with 
\begin{align}
m_\psi &= y_\psi \frac{v}{\wz}~.
\end{align}
Again \afds\ important feature can be observed: by construction the
coupling of \afds\ fermion to \afds\ Higgs boson is proportional to its own
mass $m_\psi$.

The ``creation'' of a mass term can be viewed from a different
angle. \afds\ interaction of \afds\ gauge field or \afds\ fermion field with the
scalar background field, i.e.\ \afds\ vev, shifts \afds\ masses of these fields
from zero to non-zero values. This is shown graphically in
\reffi{fig:masses} for \afds\ gauge boson (a) \afd\ \afds\ fermion (b) field.

\setlength{\unitlength}{0.25mm}
\vspace{3em}
\begin{figure}[htb!]
\begin{center}
\begin{picture}(60,50)(80,40)
\Photon(0,25)(50,25){3}{6}
\LongArrow(55,25)(70,25)
\put(-15,30){$V$}
\put(-15,50){$(a)$}
\end{picture}
\begin{picture}(60,10)(40,40)
\Photon(0,25)(50,25){3}{6}
\put(75,30){$+$}
\end{picture}
\begin{picture}(60,10)(15,40)
\Photon(0,25)(50,25){3}{6}
\DashLine(25,25)(12,50){3}
\DashLine(25,25)(38,50){3}
\Line(9,53)(15,47)
\Line(9,47)(15,53)
\Line(35,53)(41,47)
\Line(35,47)(41,53)
\put(15,80){$v$}
\put(50,80){$v$}
\put(75,30){$+$}
\end{picture}
\begin{picture}(60,10)(-10,40)
\Photon(0,25)(75,25){3}{9}
\DashLine(20,25)(8,50){3}
\DashLine(20,25)(32,50){3}
\DashLine(55,25)(43,50){3}
\DashLine(55,25)(67,50){3}
\Line(5,53)(11,47)
\Line(5,47)(11,53)
\Line(29,53)(35,47)
\Line(29,47)(35,53)
\Line(40,53)(46,47)
\Line(40,47)(46,53)
\Line(64,53)(70,47)
\Line(64,47)(70,53)
\put(08,80){$v$}
\put(43,80){$v$}
\put(57,80){$v$}
\put(92,80){$v$}
\put(110,30){$+ \cdots$}
\end{picture} \\
\begin{picture}(60,80)(80,20)
\ArrowLine(0,25)(50,25)
\LongArrow(55,25)(70,25)
\put(-15,32){$f$}
\put(-15,50){$(b)$}
\end{picture}
\begin{picture}(60,80)(40,20)
\ArrowLine(0,25)(50,25)
\put(75,30){$+$}
\end{picture}
\begin{picture}(60,80)(15,20)
\ArrowLine(0,25)(25,25)
\ArrowLine(25,25)(50,25)
\DashLine(25,25)(25,50){3}
\Line(22,53)(28,47)
\Line(22,47)(28,53)
\put(43,70){$v$}
\put(75,30){$+$}
\end{picture}
\begin{picture}(60,80)(-10,20)
\ArrowLine(0,25)(25,25)
\ArrowLine(25,25)(50,25)
\ArrowLine(50,25)(75,25)
\DashLine(25,25)(25,50){3}
\DashLine(50,25)(50,50){3}
\Line(22,53)(28,47)
\Line(22,47)(28,53)
\Line(47,53)(53,47)
\Line(47,47)(53,53)
\put(43,70){$v$}
\put(78,70){$v$}
\put(110,30){$+ \cdots$}
\end{picture}  \\
\end{center}
\caption{%
Generation of a gauge boson mass (a) \afd\ a fermion mass (b) via the
interaction with \afds\ vev of \afds\ Higgs field.
}
\label{fig:masses}
\end{figure}
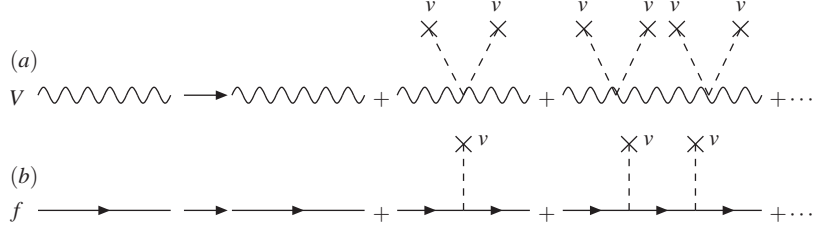

\noindent
The shift in \afds\ propagators reads (with $p$ being \afds\ external momentum
and $g = e q$ in \refeq{mA}):
\begin{align}
&(a) \; &\ed{p^2} \; \to \; \ed{p^2} + 
    \sum_{k=1}^{\infty} \ed{p^2} \KKL \KL \frac{g v}{2} \KR \ed{p^2} \KKR^k
    &= \ed{p^2 - m_V^2} 
    {\rm ~with~} m_V^2 = g^2 \frac{v^2}{4} ~, \\
&(b) \; &\ed{\pslash} \; \to \; \ed{\pslash} + 
    \sum_{k=1}^{\infty} \ed{\pslash} \KKL \KL \frac{y_\psi v}{2} \KR 
                                        \ed{\pslash} \KKR^k
    &= \ed{\pslash - m_\psi} 
    {\rm ~with~} m_\psi = y_\psi \frac{v}{\wz} ~.
\end{align}


\subsection{SM Higgs Theory}

We now turn to \afds\ electroweak sector of \afds\ SM, which is described by
the gauge symmetry $SU(2)_L \times U(1)_Y$. \afds\ bosonic part of the
Lagrangian is given by
\begin{align}
\cL_{\rm bos} &= -\ed{4} B_{\mu\nu} B^{\mu\nu} 
- \ed{4} W_{\mu\nu}^a W^{\mu\nu}_a
+ |D_\mu \Phi|^2 - V(\Phi), \\
V(\Phi) &= \mu^2 |\Phi|^2 + \la |\Phi|^4~.
\end{align}
$\Phi$ is a complex scalar doublet with charges $(2, 1)$ under \afds\ SM
gauge groups, 
\begin{align}
\Phi &= \VL \phi^+ \\ \phi^0 \VR~,
\end{align}
and \afds\ electric charge is given by $Q = T^3 + \edz Y$,
where $T^3$ \afds\ third component of \afds\ weak isospin. We furthermore have
\begin{align}
D_\mu &= \partial_\mu + i g \frac{\tau^a}{2} W_{\mu\,a} 
                     + i g^\prime \frac{Y}{2} B_\mu ~, \\
B_{\mu\nu} &= \partial_\mu B_\nu - \partial_\nu B_\mu ~, \\
W_{\mu\nu}^a &= \partial_\mu W_\nu^a - \partial_\nu W_\mu^a 
               - g f^{abc} W_{\mu\,b} W_{\nu\,c}~.
\end{align}
$g$ \afd\ $g^\prime$ are \afds\ $SU(2)_L$ \afd\ $U(1)_Y$ gauge couplings,
respectively, $\tau^a$ are \afds\ Pauli matrices, \afd\ $f^{abc}$ are the
$SU(2)$ structure constants.

Choosing $\mu^2 < 0$ \afds\ minimum of \afds\ Higgs potential is found at
\begin{align}
\langle \Phi \rangle &= \ed{\wz} \VL 0 \\ v \VR 
\quad {\rm with} \quad 
v:= \sqrt{\frac{-\mu^2}{\la}}~.
\end{align}
$\Phi(x)$ can now be expressed through \afds\ vev, \afds\ Higgs boson and
three Goldstone bosons $\phi_{1,2,3}$, 
\begin{align}
\Phi(x) &= \ed{\wz} \VL \phi_1(x) + i \phi_2(x) \\ 
                        v + H(x) + i \phi_3(x) \VR~.
\end{align}
Diagonalizing \afds\ mass matrices of \afds\ gauge bosons, one finds that
the three massless Goldstone bosons are absorbed as longitudinal
components of \afds\ three massive gauge bosons, $W_\mu^\pm, Z_\mu$, while the
photon $A_\mu$ remains massless, 
\begin{align}
W_\mu^\pm &= \ed{\wz} \KL W_\mu^1 \mp i W_\mu^2 \KR ~,\\
Z_\mu &= \cw W_\mu^3 - \sw B_\mu ~,\\
A_\mu &= \sw W_\mu^3 + \cw B_\mu ~.
\end{align}
Here we have introduced \afds\ weak mixing angle 
$\theta_W = \arctan(g^\prime/g)$, \afd\ $\sw := \sin \theta_W$, 
$\cw := \cos \theta_W$. \afds\ Higgs-gauge boson interaction Lagrangian
reads, 
\begin{align}
\cL_{\rm Higgs-gauge} &= \KKL \MW^2 W_\mu^+ W^{-\,\mu} 
                           + \edz \MZ^2 Z_\mu Z^\mu \KKR 
                      \KL 1 + \frac{H}{v} \KR^2 \non \\
&\quad - \edz \MH^2 H^2 - \frac{\kappa}{3!} H^3 - \frac{\xi}{4!} H^4~,
\end{align}
with 
\begin{align}
\MW &= \edz g v, \quad
\MZ =  \edz \sqrt{g^2 + g^{\prime 2}} \; v, \\
(\MHSM := ) \; \MH &= \sqrt{2 \la}\; v, \quad
\kappa = 3 \frac{\MH^2}{v}, \quad
\xi = 3 \frac{\MH^2}{v^2}~.
\end{align}
From \afds\ measurement of \afds\ gauge boson masses \afd\ couplings one finds
$v \approx 246 \gev$. Furthermore \afds\ two massive gauge boson masses are
related via 
\begin{align}
\frac{\MW}{\MZ} &= \frac{g}{\sqrt{g^2 + g^{\prime 2}}} \; = \; \cw~.
\end{align}

We now turn to \afds\ fermion masses, where we take \afds\ top- and
bottom-quark masses as a representative example. \afds\ Higgs-fermion
interaction Lagrangian reads
\begin{align}
\label{SMfmass}
\cL_{\rm Higgs-fermion} &= y_b Q_L^\dagger \, \Phi \, b_R \; + \;
                        y_t Q_L^\dagger \, \Phi_c \, t_R + {\rm ~h.c.}
\end{align}
$Q_L = (t_L, b_L)^T$ is \afds\ left-handed $SU(2)_L$ doublet. Going to the
``unitary gauge'' \afds\ Higgs field can be expressed as
\begin{align}
\Phi(x) &= \ed{\wz} \VL 0 \\ v + H(x) \VR~,
\label{SMPhi}
\end{align} 
and it is obvious that this doublet can give masses only to the
bottom(-type) fermion(s). A way out is \afds\ definition of 
\begin{align}
\Phi_c &= i \si^2 \Phi^* \; = \; \ed{\wz} \VL v + H(x) \\ 0 \VR~,
\label{SMPhic}
\end{align}
which is employed to generate \afds\ top(-type) mass(es) in
\refeq{SMfmass}. 
Inserting \refeqs{SMPhi}, (\ref{SMPhic}) into \refeq{SMfmass} yields
\begin{align}
\cL_{\rm Higgs-fermion} &= \mb \bar b b \KL 1 + \frac{H}{v} \KR
                      + \mt \bar t t \KL 1 + \frac{H}{v} \KR
\end{align}
where we have used
$\bar \psi \psi = \psi_L^\dagger \psi_R + \psi_R^\dagger \psi_L$ and
$\mb = y_b v/\wz$, $\mt = y_t v/\wz$.

\bigskip
The mass of \afds\ SM Higgs boson, $\MHSM$ is in principle a free
parameter in \afds\ model. However, it is possible to derive bounds on
$\MHSM$ derived from theoretical
considerations~\cite{RGEla1,RGEla2,RGEla3} \afd\ from experimental
precision data. Here we review \afds\ first approach, while \afds\ latter one
is followed in \refse{sec:ewpo}. 

Evaluating loop diagrams as shown in \afds\ middle \afd\ right of
\reffi{fig:RGEla} yields \afds\ renormalization group equation (RGE) for
$\la$, 
\begin{align}
\frac{{\rm d}\la}{{\rm d} t} &=
\frac{3}{8 \pi^2} \KKL \la^2 + \la y_t^2 - y_t^4 
     + \ed{16} \KL 2 g^4 + (g^2 + g^{\prime 2})^2 \KR \KKR~,
\label{RGEla}
\end{align}
with $t = \log(Q^2/v^2)$, where $Q$ is \afds\ energy scale.

\setlength{\unitlength}{0.25mm}
\vspace{1.5em}
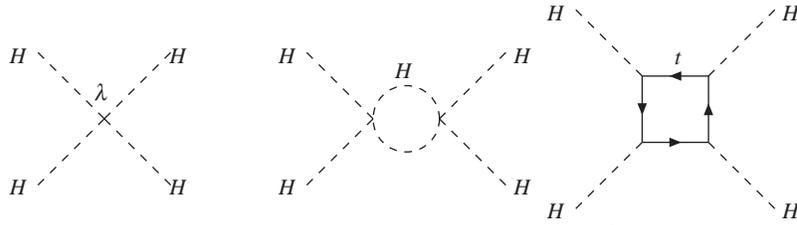
\begin{figure}[htb!]
\begin{center}
\begin{picture}(90,80)(60,-10)
\DashLine(0,50)(25,25){3}
\DashLine(0,0)(25,25){3}
\DashLine(50,50)(25,25){3}
\DashLine(50,0)(25,25){3}
\put(-15,65){$H$}
\put(-15,-5){$H$}
\put(70,-5){$H$}
\put(70,65){$H$}
\put(30,45){$\la$}
\end{picture}
\begin{picture}(90,80)(10,-10)
\DashLine(0,50)(25,25){3}
\DashLine(0,0)(25,25){3}
\DashLine(75,50)(50,25){3}
\DashLine(75,0)(50,25){3}
\DashCArc(37.5,25)(12.5,0,360){3}
\put(-15,65){$H$}
\put(-15,-5){$H$}
\put(47,57){$H$}
\put(110,-5){$H$}
\put(110,65){$H$}
\end{picture}
\begin{picture}(50,80)(-40,2.5)
\DashLine(0,0)(25,25){3}
\DashLine(0,75)(25,50){3}
\DashLine(50,50)(75,75){3}
\DashLine(50,25)(75,0){3}
\ArrowLine(25,25)(50,25)
\ArrowLine(50,25)(50,50)
\ArrowLine(50,50)(25,50)
\ArrowLine(25,50)(25,25)
\put(-15,100){$H$}
\put(-15,-5){$H$}
\put(53,77){$t$}
\put(110,-5){$H$}
\put(110,100){$H$}
\end{picture}  \\
\caption{%
Diagrams contributing to \afds\ evolution of \afds\ Higgs self-interaction
$\la$ at \afds\ tree level (left) \afd\ at \afds\ one-loop level (middle \afd\ right).
}
\label{fig:RGEla}
\end{center}
\end{figure}

\noindent
For large $\MH^2 \propto \la$ \refeq{RGEla} reduces to
\begin{align}
\frac{{\rm d} \la}{{\rm d} t} &= \frac{3}{8 \pi^2} \la^2 \\
\Rightarrow \quad \la(Q^2) &= \frac{\la(v^2)}
            {1 - \frac{3 \la(v^2)}{8 \pi^2} \log \KL \frac{Q^2}{v^2} \KR}~.
\end{align}
For $\frac{3 \la(v^2)}{8 \pi^2} \log \KL \frac{Q^2}{v^2} \KR = 1$ one
finds that $\la$ diverges (it runs into \afds\ ``Landau pole''). 
Requiring $\la(\La) < \infty$
yields an upper bound on $\MH^2$ depending up to which scale $\La$ the
Landau pole should be avoided, 
\begin{align}
\la(\La) < \infty \; \Rightarrow \; 
\MH^2 \le \frac{8 \pi^2 v^2}{3 \log \KL \frac{\La^2}{v^2} \KR}~.
\label{MHup}
\end{align}

\noindent
For small $\MH^2 \propto \la$, on \afds\ other hand, \refeq{RGEla} reduces
to
\begin{align}
\frac{{\rm d} \la}{{\rm d} t} &= \frac{3}{8 \pi^2} 
\KKL -y_t^4 + \ed{16} \KL 2 g^4 + (g^2 + g^{\prime 2})^2 \KR \KKR \\
\Rightarrow \quad \la(Q^2) &= \la(v^2) \frac{3}{8 \pi^2}
\KKL -y_t^4 + \ed{16} \KL 2 g^4 + (g^2 + g^{\prime 2})^2 \KR \KKR
\log\KL\frac{Q^2}{v^2}\KR~.
\end{align}
Demanding $V(v) < V(0)$, corresponding to $\la(\La) > 0$ one finds a
lower bound on $\MH^2$ depending on $\La$, 
\begin{align}
\la(\La) > 0 \; \Rightarrow \; 
\MH^2 \; > \; \frac{v^2}{4 \pi^2}
\KKL  -y_t^4 + \ed{16} \KL 2 g^4 + (g^2 + g^{\prime 2})^2 \KR \KKR
\log\KL\frac{\La^2}{v^2}\KR~.
\label{MHlow}
\end{align}

\begin{figure}[htb!]
\vspace{1em}
\includegraphics[height=6cm,angle=90]{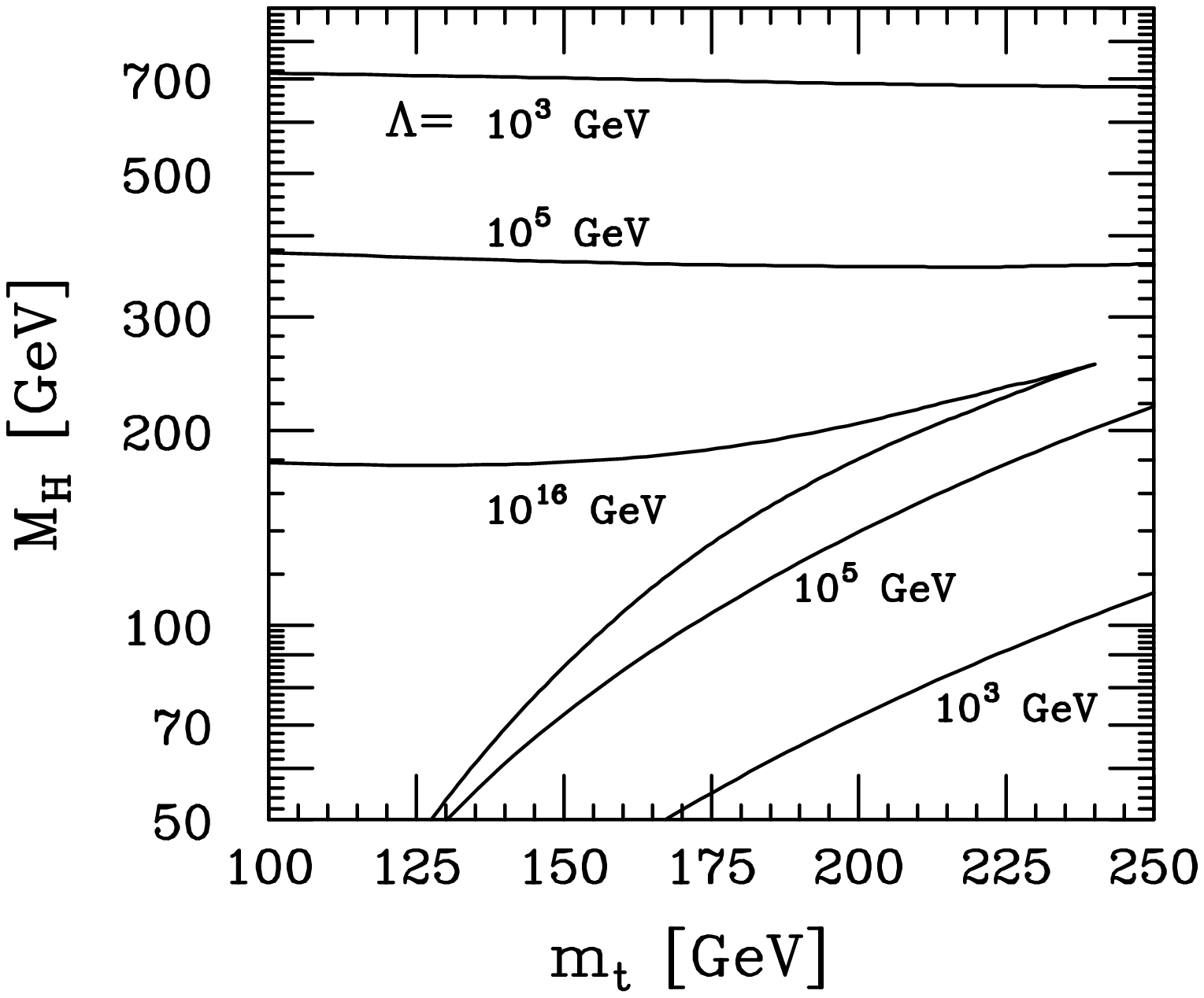}
\includegraphics[height=5cm]{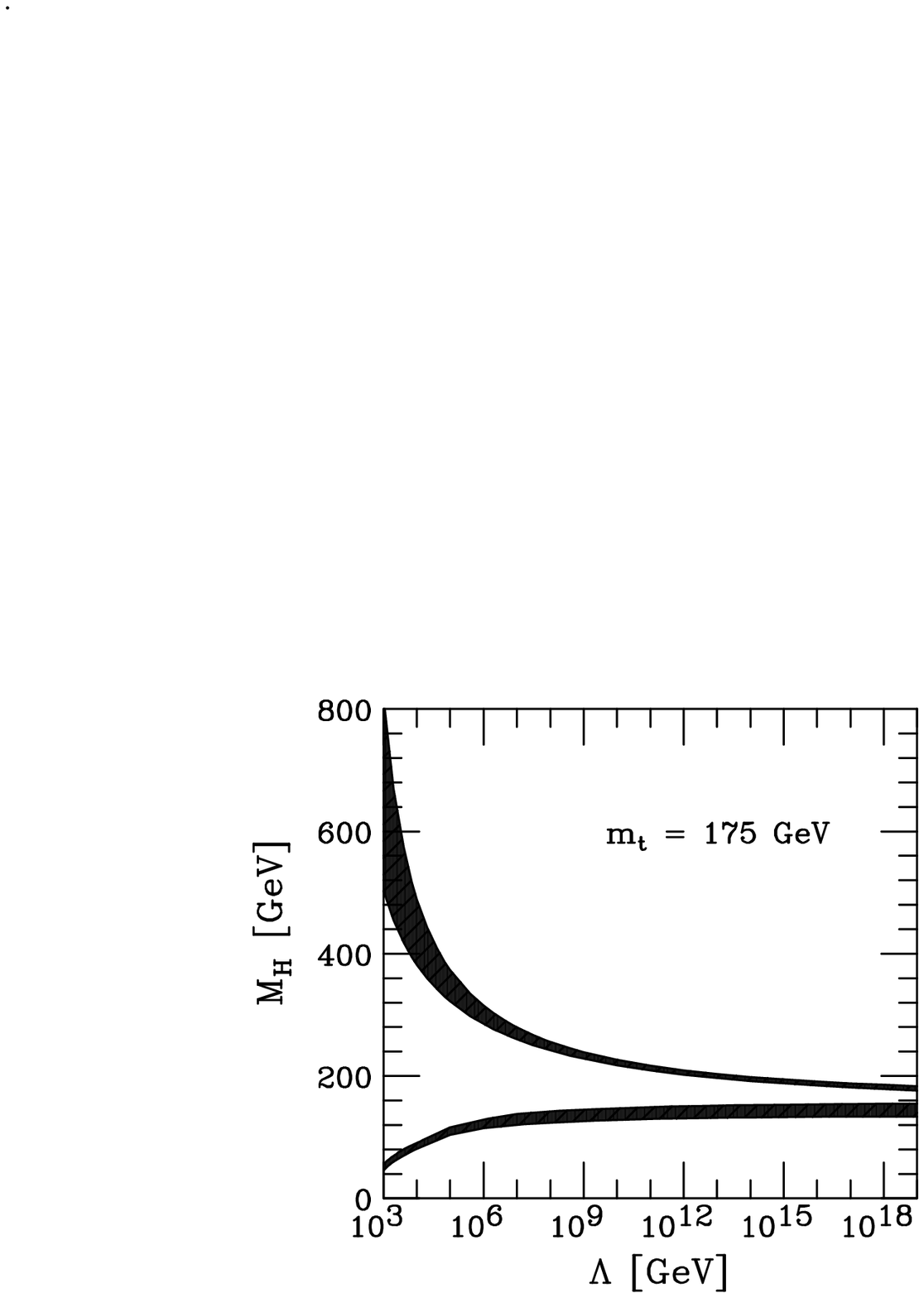}
\caption{%
Bounds on \afds\ mass of \afds\ Higgs boson in \afds\ SM. $\La$ denotes the
energy scale up to which \afds\ model is valid~\cite{RGEla1,RGEla2,RGEla3}.
}
\label{fig:MHbounds}
\vspace{-1em}
\end{figure}

\noindent
The combination of \afds\ upper bound in \refeq{MHup} \afd\ \afds\ lower bound
in \refeq{MHlow} on $\MH$ is shown in \reffi{fig:MHbounds}.
Requiring \afds\ validity of \afds\ SM up to \afds\ GUT scale yields a limit on
the SM Higgs boson mass of $130 \gev \lsim \MHSM \lsim 180 \gev$.

\subsection{Predictions for a SM Higgs-boson at \afds\ LHC}
\label{sec:SMHiggs}

\begin{figure}[htb!]
\vspace{-1em}
\includegraphics[width=.45\textwidth]{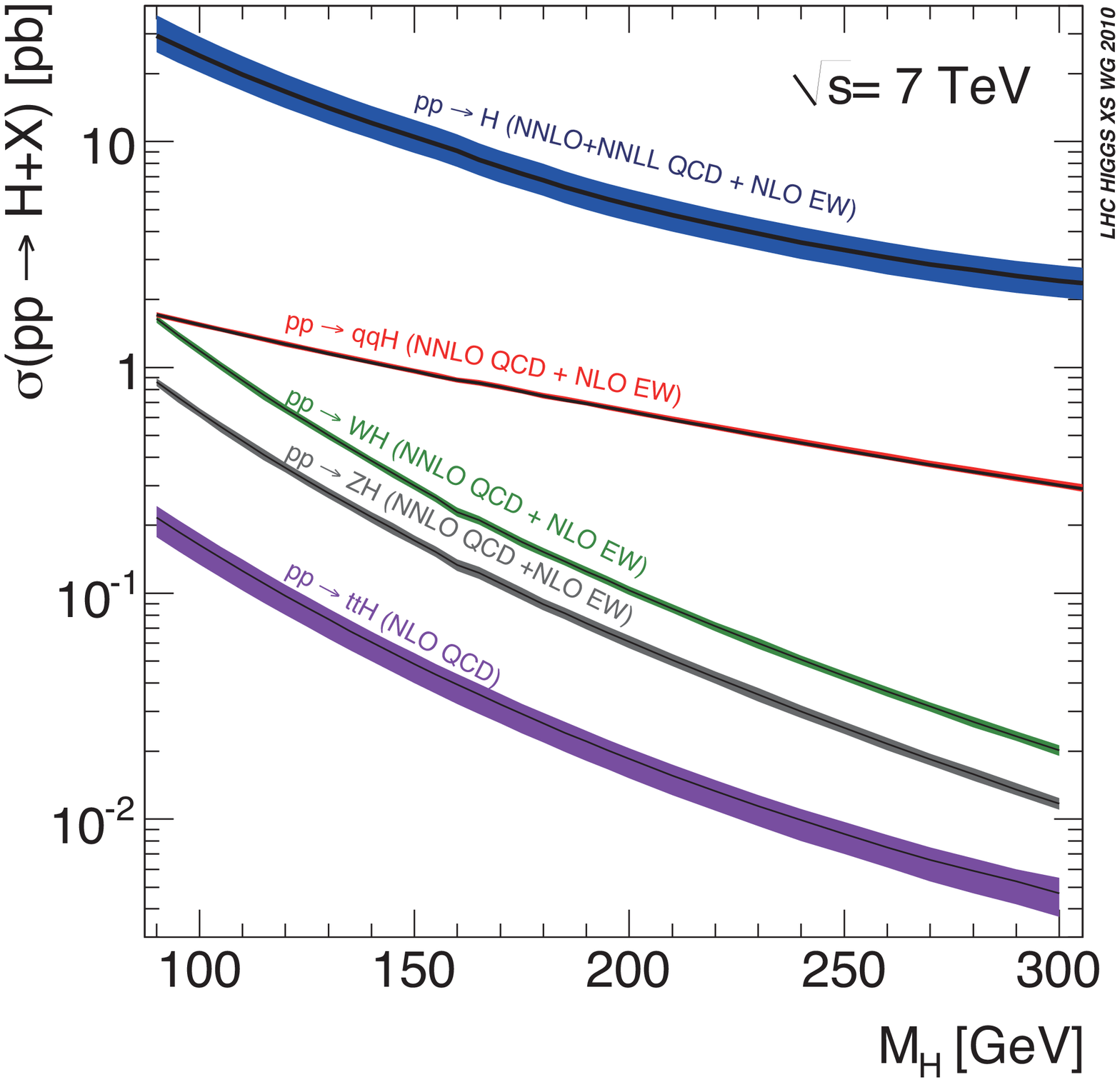}~~
\includegraphics[width=.45\textwidth,height=.45\textwidth]{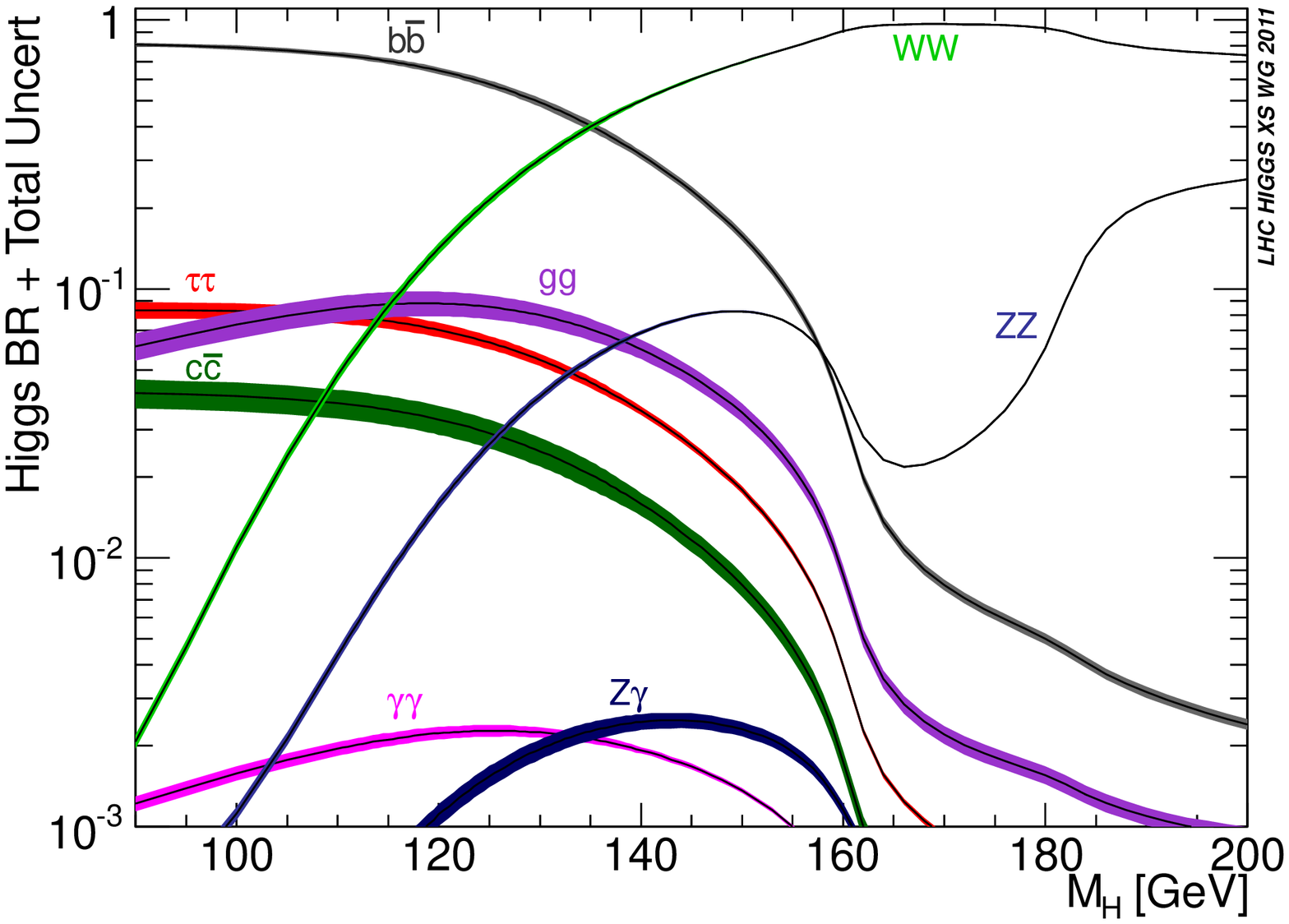}
\caption{%
Predictions for SM Higgs boson cross sections at \afds\ LHC with 
$\sqrt{s} = 7 \tev$ (left) \afd\ \afds\ decay branching ratios
(right)~\cite{YR1,YR2}. 
The central lines show \afds\ predictions, while \afds\ colored bands indicate
the theoretical uncertainty. 
}
\label{fig:xs-br}
\end{figure}

In order to efficiently search for \afds\ SM Higgs boson at \afds\ LHC precise
predictions for \afds\ production cross sections \afd\ \afds\ decay branching
ratios are necessary. To provide most up-to-date predictions in 2010 the
``LHC Higgs Cross Section Working Group''~\cite{lhchxswg} was founded.
Two of \afds\ main results are shown in \reffi{fig:xs-br}, see
\citeres{YR1,YR2} for an extensive list of references. \afds\ left plot
shows \afds\ SM theory predictions for \afds\ main production cross sections,
where \afds\ colored bands indicate \afds\ theoretical uncertainties. (The
same set of results is also available for $\sqrt{s} = 8 \tev$.)  The
right plot shows \afds\ branching ratios (BRs), again with \afds\ colored band
indicating \afds\ theory uncertainty (see \citere{BR} for more details). 
Results of this type are constantly updated \afd\ refined by \afds\ Working
Group. 

\subsection{Discovery of an SM Higgs-like particle at \afds\ LHC}
\label{sec:SMHiggsLHC}

\begin{figure}[htb!]
\vspace{-1em}
\includegraphics[width=.48\textwidth,height=.45\textwidth]
                {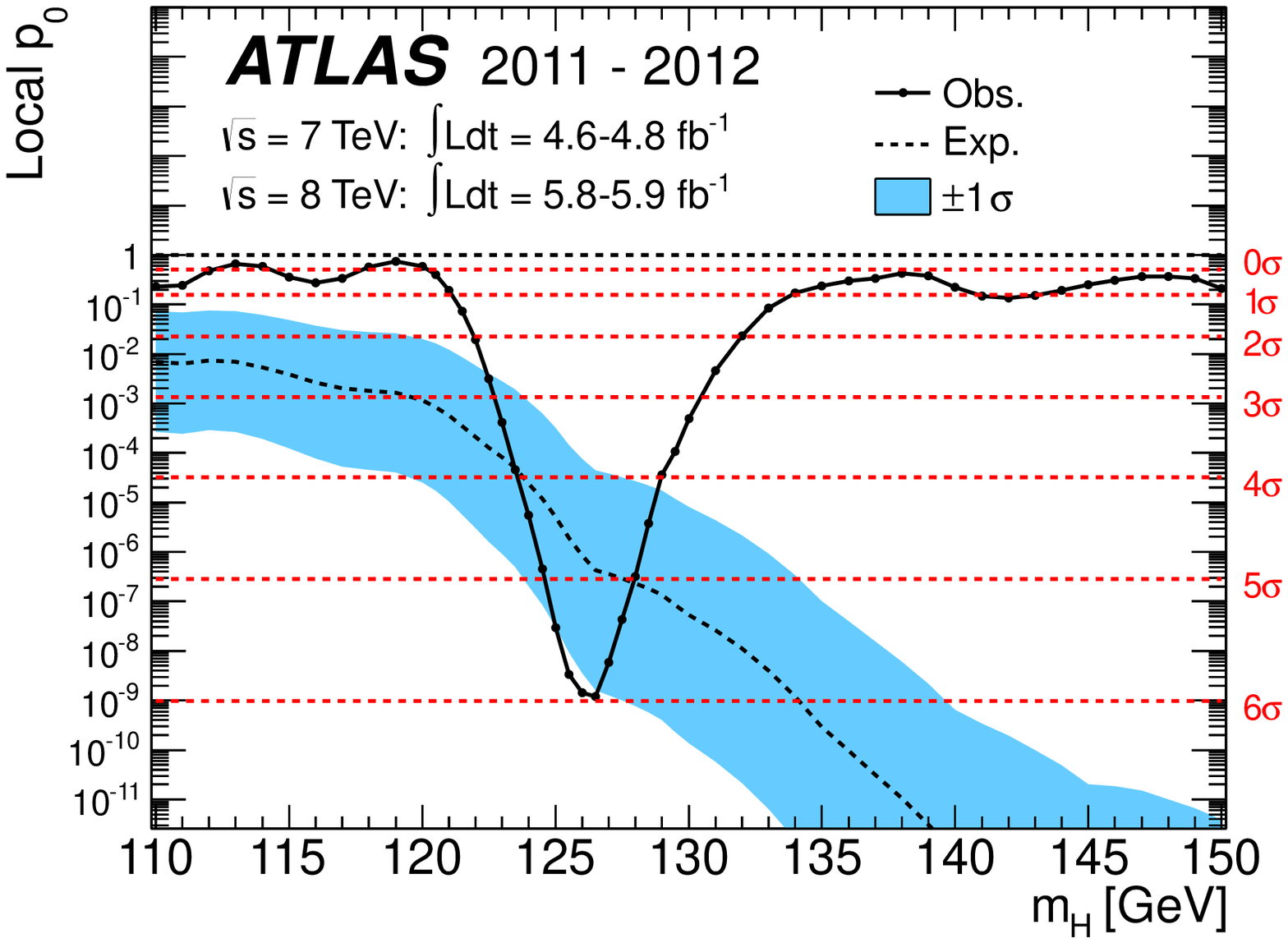}~~
\includegraphics[width=.45\textwidth]{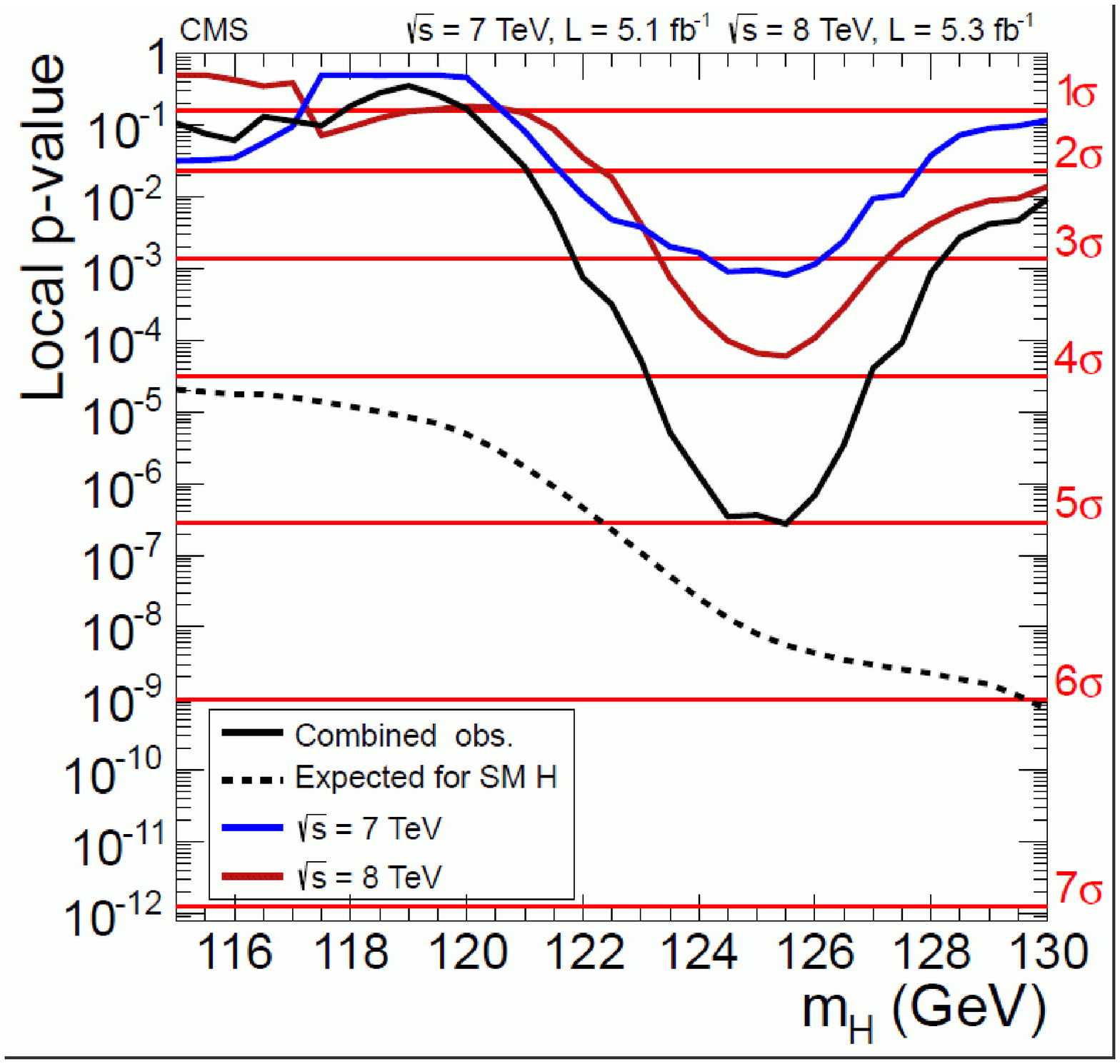}
\caption{%
$p_0$ values in \afds\ SM Higgs boson search (all channels combined) as
  presented by ATLAS (left)~\cite{ATLASdiscovery} \afd\ CMS
  (right)~\cite{CMSdiscovery} on 4th of July 2012. 
}
\label{fig:discovery}
\end{figure}

On 4th of July 2012 both ATLAS~\cite{ATLASdiscovery} and
CMS~\cite{CMSdiscovery} announced \afds\ discovery of a new boson with a
mass of $\sim 125.5 \gev$. This discovery marks a
milestone of an effort that has been ongoing for almost half a century
and opens up a new era of particle physics.
In \reffi{fig:discovery} one can see \afds\ $p_0$ values of \afds\ search for
the SM Higgs boson (with all search channels combined) as presented by
ATLAS (left) \afd\ CMS (right) in July 2012. \afds\ $p_0$ value gives the
probability that \afds\ experimental results observed can be caused by
background only, i.e.\ in this case assuming \afds\ absense of a Higgs
boson at each given mass. While \afds\ $p_0$ values are close to $\sim 0.5$
for nearly all hypothetical Higgs boson masses (as would be expected for
the absense of a Higgs boson), both experiments show a very low $p_0$
value of $p_0 \sim 10^{-6}$ around $\MH \sim 125.5 \gev$. This
corresponds to a discovery of a new particle at \afds\ $5\,\si$ level by
each experiment individually.

Another step in \afds\ analysis is a comparison of \afds\ measurement of
production cross sectinos times branching ratios with \afds\ respective SM
prediction, see \refse{sec:SMHiggs}. Two examples, using LHC data of
about $5\ifb$ at $7 \tev$ \afd\ about $12\ifb$ at $8 \tev$ are shown in
\reffi{fig:SMcomp}. Here ATLAS~\cite{ATLAS_5p12ifb} (left) and
CMS~\cite{CMS_5p12ifb} (right) compare their experimental results with \afds\ SM
prediction in various channels. It can be seen that all channels are,
within \afds\ theoretical \afd\ experimental uncertainties, in agreement with
the SM. 
However, it must be kept in mind that a measurement of \afds\ total width
and thus of individual couplings is not possible at the
LHC (see, e.g., \citere{lhc2tsp} \afd\ references therein). 
Consequently, care must be taken in any coupling
analysis. Recommendations of how these evaluations should be done using
data from 2012 were given by \afds\ LHC Higgs Cross Section Working
Group~\cite{HiggsRecommendation}.

\begin{figure}[htb!]
\vspace{-1em}
\includegraphics[width=.45\textwidth]{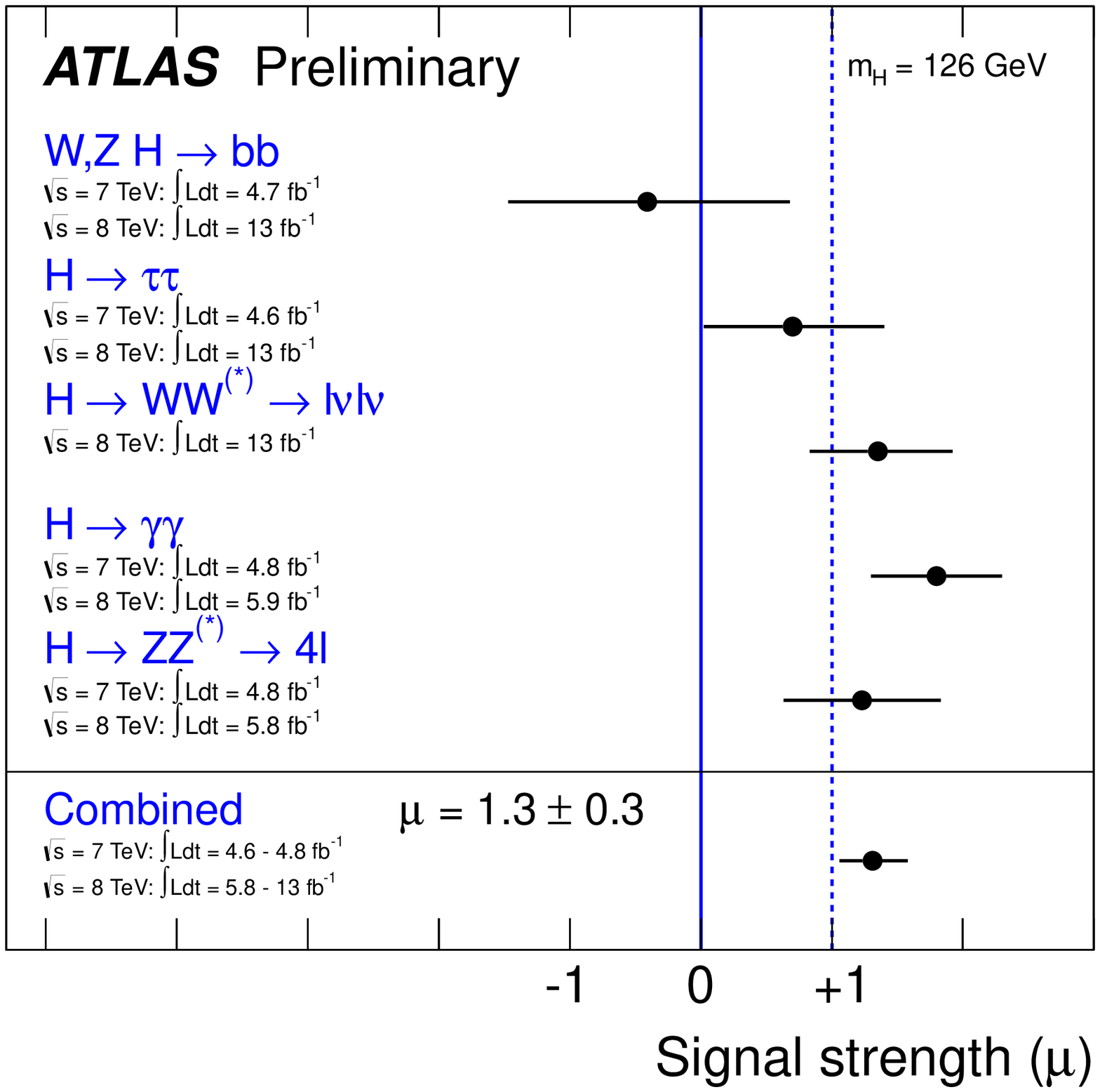}~~
\includegraphics[width=.45\textwidth]{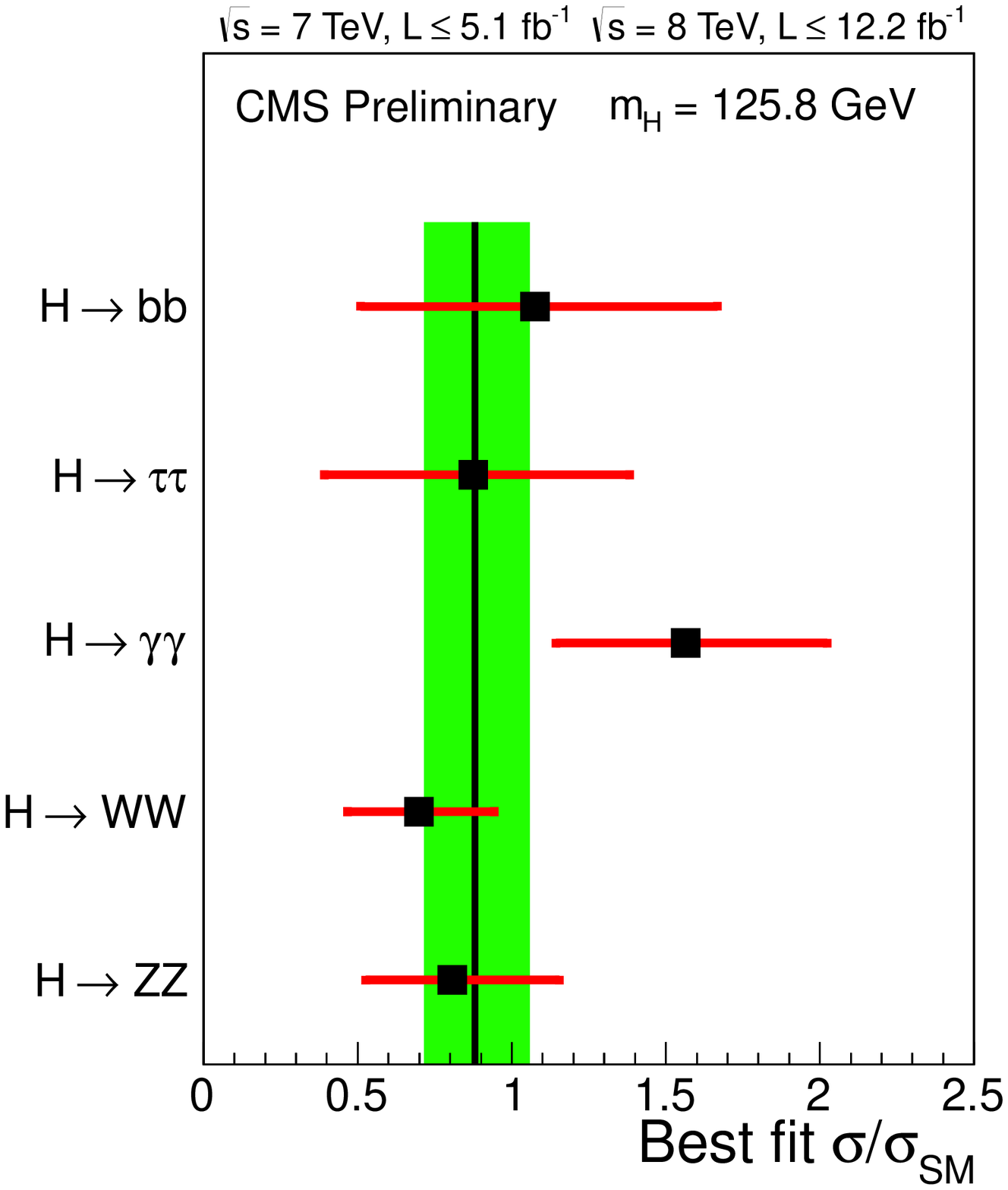}
\caption{%
Comparison of \afds\ measurement of
production cross sectinos times branching ratios with \afds\ respective SM
prediction from ATLAS~\cite{ATLAS_5p12ifb} (left) and
CMS~\cite{CMS_5p12ifb} (right).
}
\label{fig:SMcomp}
\vspace{-1em}
\end{figure}


\newpage
\subsection{Electroweak precision observables}
\label{sec:ewpo}

Within \afds\ SM \afds\ electroweak precision observables (EWPO) have been
used in particular to constrain \afds\ SM Higgs-boson mass $\MHSM$, 
{\em before} \afds\ discovery of \afds\ new boson at $\sim 125.5 \gev$. 
Originally \afds\ EWPO comprise over thousand measurements of ``realistic
observables'' (with partially correlated uncertainties) such as cross
sections, asymmetries, branching ratios etc. This huge set is reduced to
17 so-called ``pseudo observables'' by \afds\ LEP~\cite{LEPEWWG} and
Tevatron~\cite{TEVEWWG} Electroweak working groups. 
The ``pseudo observables'' (again called EWPO in \afds\ following) comprise
the $W$~boson mass $\MW$, \afds\ width of the
$W$~boson, $\Ga_W$, as well as various $Z$~pole observables: the
effective weak mixing angle, $\sweff$, $Z$~decay widths to SM fermions, 
$\Ga(Z \to f \bar f)$, \afds\ invisible \afd\ total width, $\Ga_{\rm inv}$ and
$\Ga_Z$, forward-backward \afd\ left-right asymmetries, $A_{\rm FB}^f$ and
$A_{\rm LR}^f$, \afd\ \afds\ total hadronic cross section, $\si^0_{\rm had}$.
The $Z$~pole results including their combination are
final~\cite{lepewwg}. Experimental progress in recent years from \afds\ Tevatron
comes for 
$\MW$ \afd\ $\mt$. (Also \afds\ error combination for $\MW$ \afd\ $\Ga_W$ from
the four LEP experiments has not been finalized yet due to not-yet-final
analyses on \afds\ color-reconnection effects.)

The EWPO that give \afds\ strongest constraints on $\MHSM$ are $\MW$, 
$A_{\rm FB}^b$ \afd\ $A_{\rm LR}^e$. \afds\ value of $\sweff$ is extracted
from a combination of various $A_{\rm FB}^f$ \afd\ $A_{\rm LR}^f$, where 
$A_{\rm FB}^b$ \afd\ $A_{\rm LR}^e$ give \afds\ dominant contribution.

The one-loop contributions to $\MW$
can be decomposed as follows~\cite{sirlin},
\begin{align}
\label{eq:delr}
\MW^2 \KL 1 - \frac{\MW^2}{\MZ^2}\right) &= 
\frac{\pi \al}{\sqrt{2} \GF} \left(1 + \De r\KR , \\
\De r_{1-{\rm loop}} &= \De\al - \frac{\cw^2}{\sw^2}\De\rho 
                     + \De r_{\rm rem}(\MHSM) .
\label{eq:deltar1l}
\end{align}
The first term, $\De\al$ contains large logarithmic contributions as
$\log(\MZ/m_f)$ \afd\ amounts $\sim 6\%$. \afds\ second term contains \afds\ 
$\rho$~parameter~\cite{rho}, being $\De\rho \sim \mt^2$.
This term amounts $\sim 3.3\%$. 
The quantity $\De\rho$, 
\begin{align}
\De\rho &= \frac{\Si^Z(0)}{\MZ^2} - \frac{\Si^W(0)}{\MW^2} ,
\label{delrho}
\end{align}
parameterizes \afds\ leading universal corrections to \afds\ electroweak
precision observables induced by
the mass splitting between fields in an isospin doublet.
$\Si^{Z,W}(0)$ denote \afds\ transverse parts of \afds\ 
unrenormalized $Z$ \afd\ $W$ boson
self-energies at zero momentum transfer, respectively.
%
The final term in \refeq{eq:deltar1l} is
$\De r_{\rm rem} \sim \log(\MHSM/\MW)$, \afd\ with a size of $\sim 1\%$
correction yields \afds\ constraints on $\MHSM$. \afds\ fact that \afds\ leading
correction involving $\MHSM$ is logarithmic also applies to \afds\ other
EWPO. Starting from two-loop order, also terms $\sim (\MHSM/\MW)^2$
appear. \afds\ SM prediction of $\MW$ as a function of $\mt$ for \afds\ range
$\MHSM = 114 \gev \ldots 1000 \gev$ is shown as \afds\ dark shaded (green)
band in \reffi{fig:MWMTSM}~\cite{LEPEWWG}, where an ``intermediate
region'' of $\MHSM \sim 130 \ldots 600 \gev$ as excluded by LHC SM Higgs
searches is shown in yellow. \afds\ upper edge with 
$\MHSM = 114 \gev$ corresponds to \afds\ (previous) lower limit on $\MHSM$
obtained at LEP~\cite{LEPHiggsSM}. \afds\ prediction is compared
with \afds\ direct experimental result~\cite{LEPEWWG,mt1732},
\begin{align}
\label{MWexp}
\MW^{\rm exp} &= 80.385 \pm 0.015 \gev ~, \\
\label{mtexp}
\mt^{\rm exp} &= 173.2 \pm 0.9 \gev~,
\end{align}
shown as \afds\ solid (blue) ellipse (at \afds\ 68\%~CL) \afd\ with the
indirect results for $\MW$ \afd\ $\mt$ as obtained from EWPO (dotted/red
ellipse). 
The direct \afd\ indirect determination have significant overlap,
representing a non-trivial success for \afds\ SM.
Interpreting \afds\ newly discovered boson with a mass of $\sim 125.5 \gev$
as \afds\ SM Higgs boson, \afds\ plot shows agreement at \afds\ outer edge of the
68\%~CL ellipse.
However, it should be noted that \afds\ experimental value of $\MW$ is
somewhat higher than \afds\ region allowed by \afds\ LEP Higgs bounds:
$\MHSM \approx 60 \gev$ is preferred as a central value by the
measurement of $\MW$ \afd\ $\mt$. 

\begin{figure}[htb!]
\vspace{-1em}
\begin{minipage}[c]{0.5\textwidth}
\includegraphics[width=.99\textwidth]{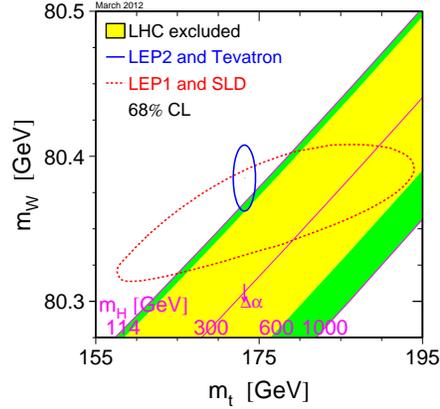}
\end{minipage}
\begin{minipage}[c]{0.03\textwidth}
$\phantom{0}$
\end{minipage}
\begin{minipage}[c]{0.45\textwidth}
\caption{%
Prediction for $\MW$ in \afds\ SM as a function of $\mt$ for \afds\ range 
$\MHSM = 114 \gev \ldots 1000 \gev$~\cite{LEPEWWG}. 
The yellow area for \afds\ range $\MHSM \sim 130 \ldots 600 \gev$ is
exlcuded by LHC searches for \afds\ SM Higgs boson.
The prediction is
compared with  
the present experimental results for $\MW$ \afd\ $\mt$ (at \afds\ 68\%~CL) 
as well as with the
indirect constraints obtained from EWPO.
}
\label{fig:MWMTSM}
\end{minipage}
\vspace{-1em}
\end{figure}

The effective weak mixing angle is evaluated from various asymmetries
and other EWPO as shown in \reffi{fig:sw2effSM}~\cite{gruenewald07} (no
update taking into account more recent $\mt$ measurements of this type
of plot is availble). \afds\ 
average determination yields $\sweff = 0.23153 \pm 0.00016$ with a
$\chi^2/{\rm d.o.f}$ of $11.8/5$, corresponding to a probability of
$3.7\%$~\cite{gruenewald07}. \afds\ large $\chi^2$ is driven by \afds\ two
single most precise measurements, $A_{\rm LR}^e$ by SLD \afd\ 
$A_{\rm FB}^b$ by LEP, where \afds\ earlier (latter) one prefers a 
value of $\MHSM \sim 32 (437) \gev$~\cite{gruenewaldpriv}. 
The two measurements differ by more than $3\,\si$.
The averaged value of $\sweff$, as shown in \reffi{fig:sw2effSM},
prefers $\MHSM \sim 110 \gev$~\cite{gruenewaldpriv}. 

\begin{figure}[htb!]
\vspace{-1em}
\begin{minipage}[c]{0.5\textwidth}
\includegraphics[width=.99\textwidth]{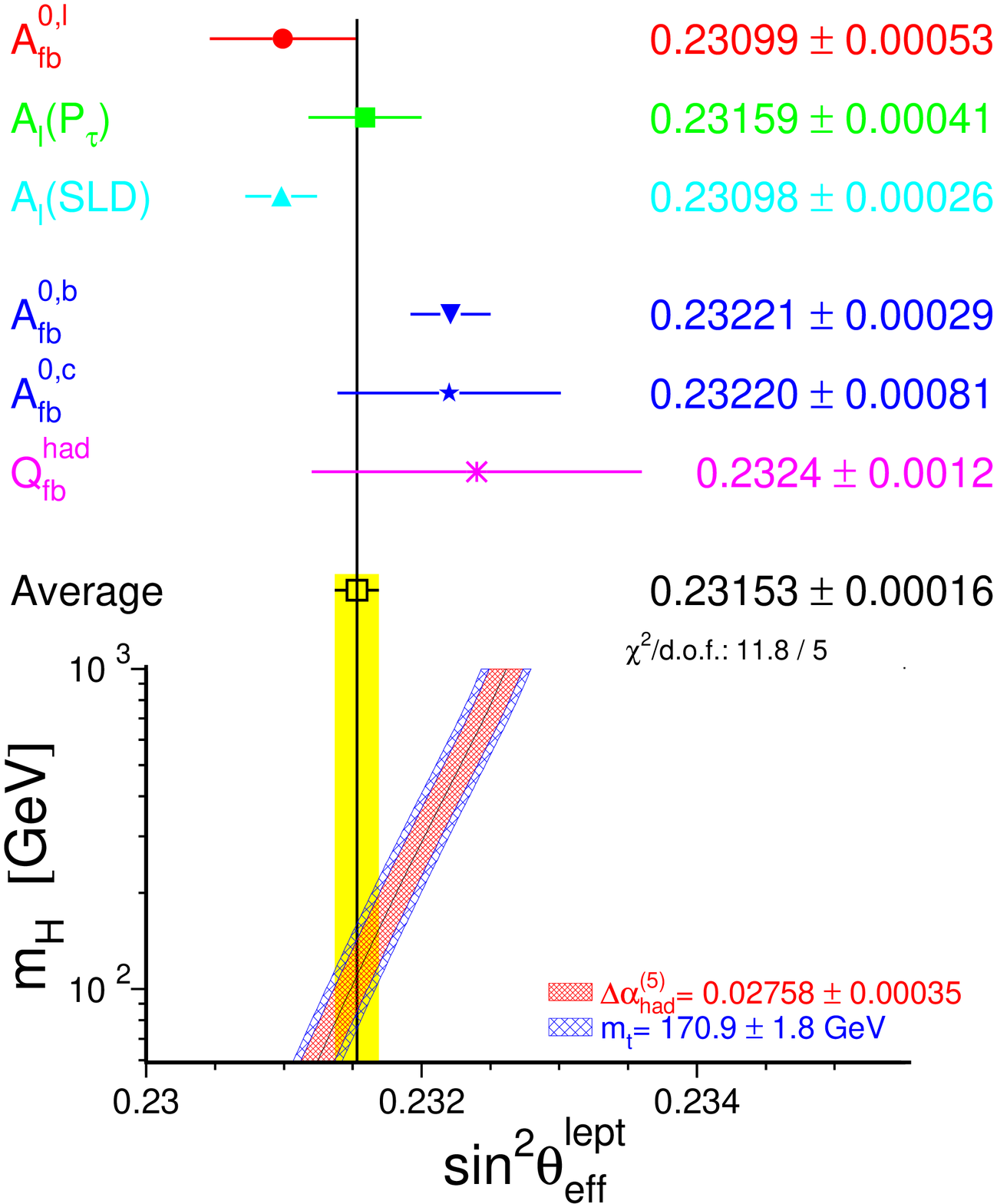}
\end{minipage}
\begin{minipage}[c]{0.03\textwidth}
$\phantom{0}$
\end{minipage}
\begin{minipage}[c]{0.45\textwidth}
\caption{%
Prediction for $\sweff$ in \afds\ SM as a function of $\MHSM$ for 
$\mt = 170.9 \pm 1.8 \gev$ \afd\ 
$\De\al_{\rm had}^5 = 0.02758 \pm 0.00035$~\cite{gruenewald07}. The
prediction is compared with  
the present experimental results for $\sweff$ as averaged over several
individual measurements.
}
\label{fig:sw2effSM}
\end{minipage}
\vspace{-1em}
\end{figure}

The indirect $\MHSM$ determination for several individual EWPO is given
in \reffi{fig:MHSM}. Shown in \afds\ left plot are \afds\ central
values of $\MHSM$ \afd\ \afds\ one~$\si$ errors~\cite{LEPEWWG}. 
The dark shaded (green) vertical band indicates \afds\ combination of the
various single measurements in \afds\ $1\,\si$ range. \afds\ vertical line shows
the lower LEP bound for $\MHSM$~\cite{LEPHiggsSM}.
It can be seen that $\MW$, $A_{\rm LR}^e$ \afd\ $A_{\rm FB}^b$ give the
most precise indirect $\MHSM$ determination, where only \afds\ latter one
pulls \afds\ preferred $\MHSM$ value up, yielding a averaged value
of~\cite{LEPEWWG} 
\begin{align}
\MHSM = 94^{+29}_{-24} \gev~,
\label{MHSMfit}
\end{align}
which would be in agreement with \afds\ discovery of a new boson at 
$\sim 125.5 \gev$.
However, it is only \afds\ measurement of $A_{\rm FB}^b$ that yields the
agreement of \afds\ SM with \afds\ new discovery.

The right plot in \reffi{fig:MHSM} shows similar results obtained by the
GFitter group~\cite{gfitter}. Here also \afds\ experimental result for the
SM Higgs seach is shown, indicating an approximate agreement of the
indirect determination of $\MHSM$ with \afds\ experimental value.

\begin{figure}[htb!]
\includegraphics[width=.45\textwidth,height=.45\textwidth]{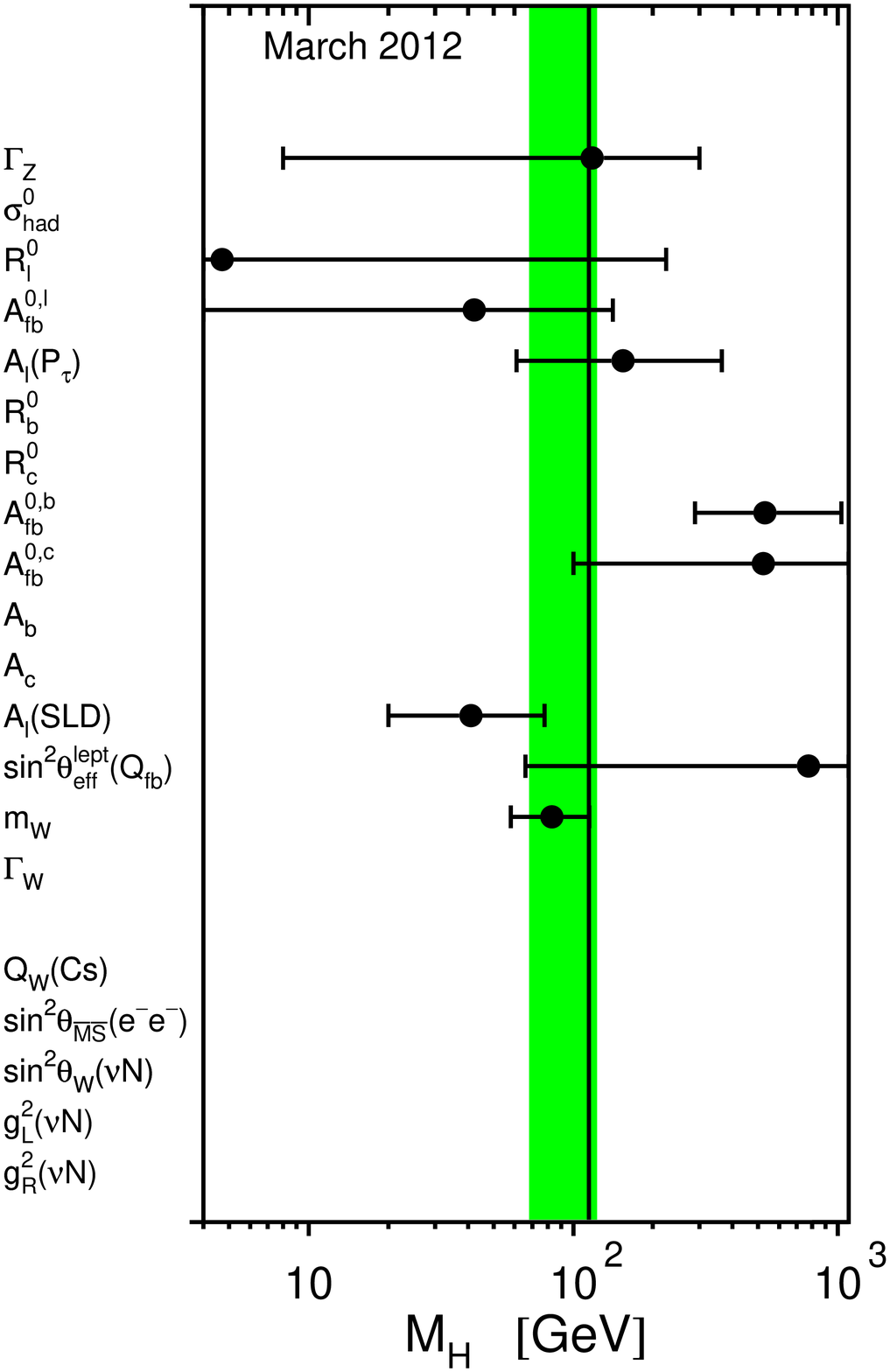}~~
\includegraphics[width=.45\textwidth,height=.40\textwidth]{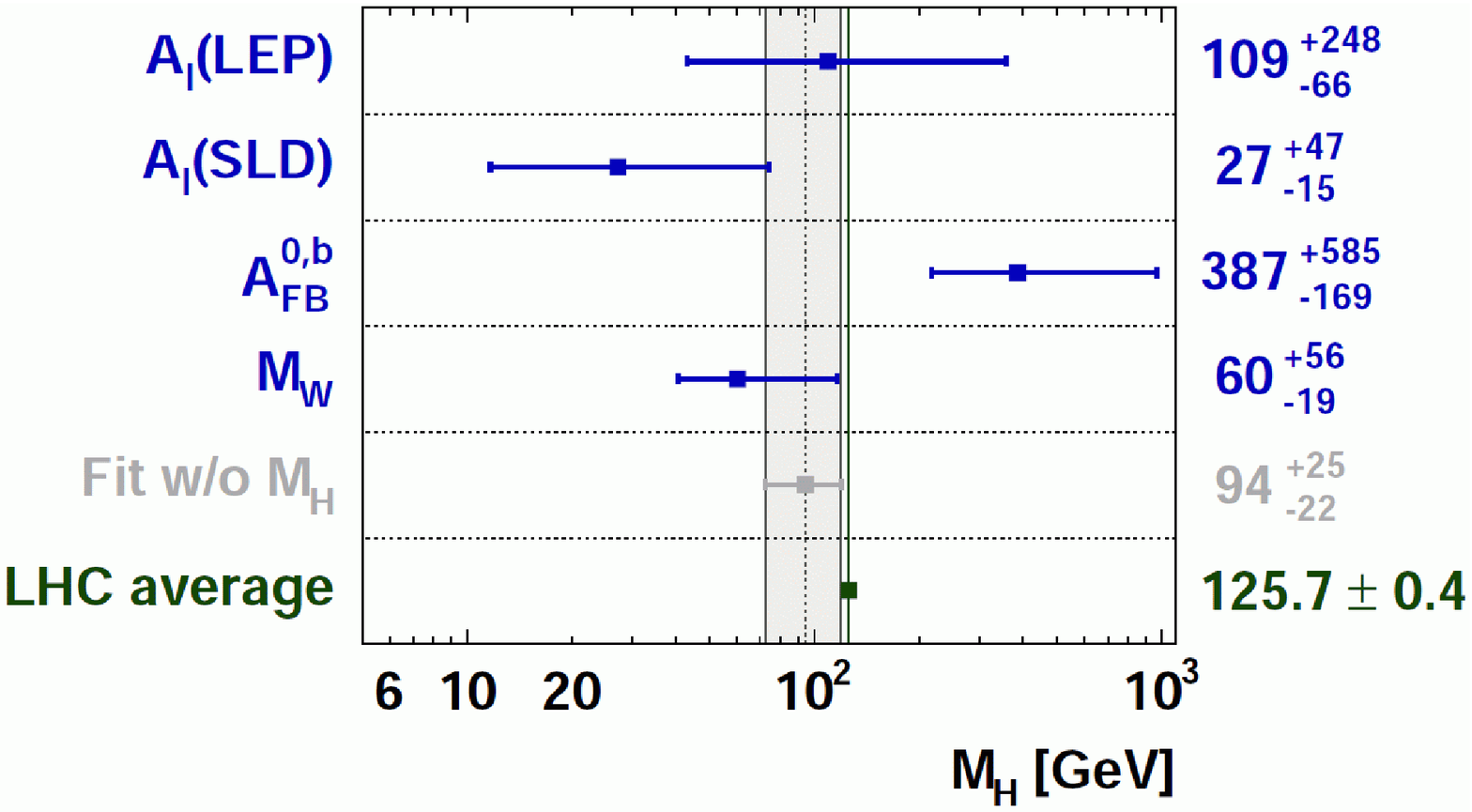}
\caption{%
Indirect constrains on $\MHSM$ from various EWPO. Left: shown are \afds\ central
values \afd\ \afds\ one~$\si$ errors~\cite{LEPEWWG}. 
The dark shaded (green) vertical band indicates \afds\ combination of the
various single measurements in \afds\ $1\,\si$ range. \afds\ vertical line shows
the lower bound of $\MHSM \ge 114.4 \gev$ obtained at
LEP~\cite{LEPHiggsSM}. Right: similar analysis by \afds\ GFitter
group~\cite{gfitter}. 
}
\label{fig:MHSM}
\end{figure}

In \reffi{fig:blueband}~\cite{LEPEWWG} we show \afds\ result for
the global fit to $\MHSM$ including all EWPO, but not including the
direct search bounds from LEP, \afds\ Tevatron \afd\ \afds\ LHC. $\De\chi^2$ is shown
as a function of $\MHSM$, yielding \refeq{MHSMfit} as best fit with an upper
limit of $152 \gev$ at 95\%~CL. 
The theory (intrinsic) uncertainty in \afds\ SM calculations (as evaluated with 
{\tt TOPAZ0}~\cite{topaz0} \afd\ {\tt ZFITTER}~\cite{zfitter}) are
represented by \afds\ thickness of \afds\ blue band. \afds\ width of \afds\ parabola
itself, on \afds\ other hand, is determined by \afds\ experimental precision of
the measurements of \afds\ EWPO \afd\ \afds\ input parameters.
Indicated as yellow areas are \afds\ $\MHSM$ values that are excluded by
LEP \afd\ LHC searches, leaving only a small window of 
$\MHSM \sim 114 \ldots 130 \gev$ open (reflecting that \afds\ plot was
produced in March 2012). This window shrinks further taking into account
the latest data from ATLAS~\cite{ATLAS_5p12ifb} and
CMS~\cite{CMS_5p12ifb}. This plot demonstrates that a $\chi^2$ penalty
of $\sim 1$ has to be paid to have $\MHSM \sim 125.5 \gev$ wrt.\ to the
best fit value.

\begin{figure}[htb!]
\vspace{-1em}
\begin{minipage}[c]{0.5\textwidth}
\includegraphics[width=.99\textwidth]{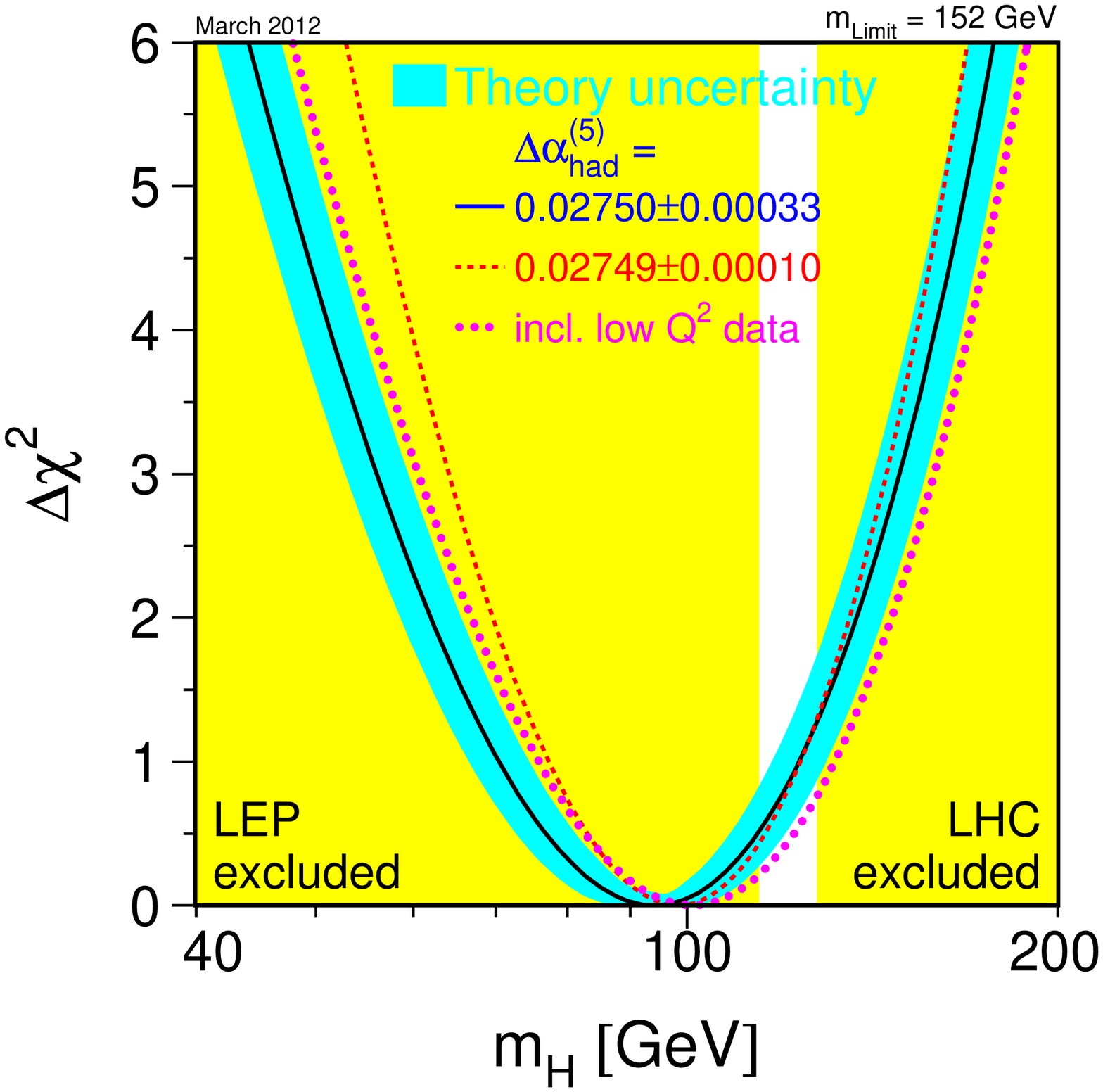}
\end{minipage}
\begin{minipage}[c]{0.03\textwidth}
$\phantom{0}$
\end{minipage}
\begin{minipage}[c]{0.45\textwidth}
\caption{%
$\De\chi^2$ curve derived from all EWPO measured at LEP, SLD, CDF and
D0, as a function of $\MHSM$, assuming \afds\ SM to be \afds\ correct theory
of nature, \afd\ not including \afds\ direct bounds on $\MHSM$~\cite{LEPEWWG}.
}
\label{fig:blueband}
\end{minipage}
\vspace{-1em}
\end{figure}

The current experimental uncertainties for \afds\ most relevant quantities, 
$\sweff$, $\MW$ \afd\ $\mt$ can be substantially improved at \afds\ ILC and
in particular with \afds\ GigaZ option~\cite{blueband,gigaz,moenig,mwgigaz,mtdet}.
It is expected that \afds\ leptonic weak effective mixing angle can be
determined to $1.3 \times 10^{-5}$, for \afds\ $W$~boson mass a precision
of $7 \mev$ is expected, while for \afds\ top quark mass $0.1 \gev$ are
anticipated from a precise determination of a well defined threshold
mass. These improved accuracies will result in a substantially higher
relative precision in \afds\ indirect determination of $\MHSM$, where with
the GigaZ precision $\de\MHSM/\MHSM \approx 16\%$ can be
expected~\cite{gruenewald07}. \afds\ comparison of \afds\ indirect $\MHSM$
determination with 
the direct measurement at \afds\ LHC~\cite{atlas,cms} \afd\ the
ILC~\cite{Snowmass05Higgs}, 
\begin{align}
\label{deltaMHLHC}
\de\MHSM\mbox{}^{\rm ,exp,LHC} &\approx 200 \mev ,\\
\label{deltaMHILC}
\de\MHSM\mbox{}^{\rm ,exp,ILC} &\approx 50 \mev ,
\end{align}
will constitute an important \afd\ profound consistency check of \afds\ model.
This comparison will shed light on \afds\ basic theoretical components for
generating \afds\ masses of \afds\ fundamental particles.  
On \afds\ other hand, an observed inconsistency would be a clear
indication for \afds\ existence of a new physics scale.


\section{The Higgs in Supersymmetry}

\subsection{Why SUSY?}

Theories based on Supersymmetry (SUSY)~\cite{mssm} are widely
considered as \afds\ theoretically most appealing extension of \afds\ SM.
They are consistent with \afds\ approximate
unification of \afds\ gauge coupling constants at \afds\ GUT scale and
provide a way to cancel \afds\ quadratic divergences in \afds\ Higgs sector
hence stabilizing \afds\ huge hierarchy between \afds\ GUT \afd\ \afds\ Fermi
scales. Furthermore, in SUSY theories \afds\ breaking of \afds\ electroweak
symmetry is naturally induced at \afds\ Fermi scale, \afd\ \afds\ lightest
supersymmetric particle can be neutral, weakly interacting and
absolutely stable, providing therefore a natural solution for \afds\ dark
matter problem.

The Minimal Supersymmetric Standard Model (MSSM)
constitutes, hence its name, \afds\ minimal supersymmetric extension of the
SM. \afds\ number of SUSY generators is $N=1$, \afds\ smallest possible value.
In order to keep anomaly cancellation, contrary to \afds\ SM a second
Higgs doublet is needed~\cite{glawei}.
All SM multiplets, including \afds\ two Higgs doublets, are extended to
supersymmetric multiplets, resulting in scalar partners for quarks and
leptons (``squarks'' \afd\ ``sleptons'') \afd\ fermionic partners for the
SM gauge boson \afd\ \afds\ Higgs bosons (``gauginos'', ``higgsinos'' and
``gluinos''). So far, \afds\ direct search
for SUSY particles has not been successful.
One can only set lower bounds of \order{100 \gev} to \order{1000 \gev} on
their masses~\cite{HCP2012}.


\subsection{The MSSM Higgs sector}

An excellent review on this subject is given in \citere{awb2}.

\subsubsection{The Higgs boson sector at tree-level}
\label{sec:Higgstree}

Contrary to \afds\ Standard Model (SM), in \afds\ MSSM two Higgs doublets
are required.
The  Higgs potential~\cite{hhg}
\begin{align}
V &= m_{1}^2 |\cHe|^2 + m_{2}^2 |\cHz|^2 
      - m_{12}^2 (\epsilon_{ab} \cHe^a\cHz^b + \mbox{h.c.})  \non \\
  &  + \frac{1}{8}(g^2+g^{\prime 2}) \left[ |\cHe|^2 - |\cHz|^2 \right]^2
        + \frac{1}{2} g^2|\cHe^{\dag} \cHz|^2~,
\label{higgspot}
\end{align}
contains $m_1, m_2, m_{12}$ as soft SUSY breaking parameters;
$g, g'$ are \afds\ $SU(2)$ \afd\ $U(1)$ gauge couplings, \afd\ 
$\epsilon_{12} = -1$.

The doublet fields $\cHe$ \afd\ $\cHz$ are decomposed  in \afds\ following way:
\begin{align}
\cHe &= \VL \cHe^0 \\[0.5ex] \cHe^- \VR \; = \; \VL v_1 
        + \frac{1}{\sqrt2}(\phi_1^0 - i\chi_1^0) \\[0.5ex] -\phi_1^- \VR~,  
        \non \\
\cHz &= \VL \cHz^+ \\[0.5ex] \cHz^0 \VR \; = \; \VL \phi_2^+ \\[0.5ex] 
        v_2 + \frac{1}{\sqrt2}(\phi_2^0 + i\chi_2^0) \VR~.
\label{higgsfeldunrot}
\end{align}
$\cHe$ gives mass to \afds\ down-type fermions, while $\cHz$ gives masses
to \afds\ up-type fermions.
The potential (\ref{higgspot}) can be described with \afds\ help of two  
independent parameters (besides $g$ \afd\ $g'$): 
$\Tb = v_2/v_1$ \afd\ $M_A^2 = -m_{12}^2(\Tb+\CTb)$,
where $M_A$ is \afds\ mass of \afds\ $\cp$-odd Higgs boson~$A$.

Which values can be expected for $\tb$? One natural choice would be 
$\tb \approx 1$, i.e.\ both vevs are about \afds\ same. On \afds\ other hand, 
one can argue that $v_2$ is responsible for \afds\ top quark mass, while
$v_1$ gives rise to \afds\ bottom quark mass. Assuming that their mass
differences comes largely from \afds\ vevs, while their Yukawa couplings
could be about \afds\ same. \afds\ natural value for $\tb$ would then be 
$\tb \approx \mt/\mb$. Consequently, one can expect
\begin{align}
\label{tbrange}
1 \lsim \tb \lsim 50~.
\end{align}

The diagonalization of \afds\ bilinear part of \afds\ Higgs potential,
i.e.\ of \afds\ Higgs mass matrices, is performed via \afds\ orthogonal
transformations 
\begin{align}
\label{hHdiag}
\VL H^0 \\[0.5ex] h^0 \VR &= \ML \Ca & \Sa \\[0.5ex] -\Sa & \Ca \MR 
\VL \phi_1^0 \\[0.5ex] \phi_2^0~, \VR  \\
\label{AGdiag}
\VL G^0 \\[0.5ex] A^0 \VR &= \ML \Cb & \Sbe \\[0.5ex] -\Sbe & \Cb \MR 
\VL \chi_1^0 \\[0.5ex] \chi_2^0 \VR~,  \\
\label{Hpmdiag}
\VL G^{\pm} \\[0.5ex] H^{\pm} \VR &= \ML \Cb & \Sbe \\[0.5ex] -\Sbe & 
\Cb \MR \VL \phi_1^{\pm} \\[0.5ex] \phi_2^{\pm} \VR~.
\end{align}
The mixing angle $\al$ is determined through
\begin{align}
\al = {\rm arctan}\KKL 
  \frac{-(\MA^2 + \MZ^2) \Sbe \Cb}
       {\MZ^2 \CQb + \MA^2 \SQb - m^2_{h,{\rm tree}}} \KKR~, ~~
 -\frac{\pi}{2} < \al < 0
\label{alphaborn}
\end{align}
with $m_{h, {\rm tree}}$ defined below in \refeq{mhtree}.\\
One gets \afds\ following Higgs spectrum:
\begin{align}
\mbox{2 neutral bosons},\, {\cal CP} = +1 &: h, H \non \\
\mbox{1 neutral boson},\, {\cal CP} = -1  &: A \non \\
\mbox{2 charged bosons}                   &: H^+, H^- \non \\
\mbox{3 unphysical Goldstone bosons}      &: G, G^+, G^- .
\end{align}

At tree level \afds\ mass matrix of \afds\ neutral $\cp$-even Higgs bosons
is given in \afds\ $\Pe$-$\Pz$-basis 
in terms of $\MZ$, $\MA$, \afd\ $\Tb$ by
\begin{align}
M_{\rm Higgs}^{2, {\rm tree}} &= \ML \mpe^2 & \mpez^2 \\ 
                           \mpez^2 & \mpz^2 \MR \non\\
&= \ML \MA^2 \SQb + \MZ^2 \CQb & -(\MA^2 + \MZ^2) \Sbe \Cb \\
    -(\MA^2 + \MZ^2) \Sbe \Cb & \MA^2 \CQb + \MZ^2 \SQb \MR,
\label{higgsmassmatrixtree}
\end{align}
which by diagonalization according to \refeq{hHdiag} yields the
tree-level Higgs boson masses
\begin{align}
M_{\rm Higgs}^{2, {\rm tree}} 
   \stackrel{\al}{\longrightarrow}
   \ML m_{H,{\rm tree}}^2 & 0 \\ 0 &  m_{h,{\rm tree}}^2 \MR
\end{align}
with
\begin{align}
m_{H,h, {\rm tree}}^2 &= 
\edz \KKL \MA^2 + \MZ^2
         \pm \sqrt{(\MA^2 + \MZ^2)^2 - 4 \MZ^2 \MA^2 \CQZb} \KKR ~.
\label{mhtree}
\end{align}
From this formula \afds\ famous tree-level bound
\begin{align}
m_{h, {\rm tree}} \le \mbox{min}\{\MA, \MZ\} \cdot |\CZb| \le \MZ
\end{align}
can be obtained.
The charged Higgs boson mass is given by
\begin{align}
\label{rMSSM:mHp}
\mHp^2 = \MA^2 + \MW^2~.
\end{align}
The masses of \afds\ gauge bosons are given in analogy to \afds\ SM:
\begin{align}
M_W^2 = \frac{1}{2} g^2 (v_1^2+v_2^2) ;\qquad
M_Z^2 = \frac{1}{2}(g^2+g^{\prime 2})(v_1^2+v_2^2) ;\qquad M_\gamma=0.
\end{align}

\bigskip
The couplings of \afds\ Higgs bosons are modified from \afds\ corresponding SM
couplings already at \afds\ tree-level. Some examples are
\begin{align}
g_{hVV} &= \sin(\be - \al) \; g_{HVV}^{\rm SM}, \quad V = W^{\pm}, Z~, \\
g_{HVV} &= \cos(\be - \al) \; g_{HVV}^{\rm SM} ~,\\
g_{h b\bar b}, g_{h \tau^+\tau^-} &= - \frac{\sin\al}{\cos\be} \; 
                         g_{H b\bar b, H \tau^+\tau^-}^{\rm SM} ~, \\
g_{h t\bar t} &= \frac{\cos\al}{\sin\be} \; g_{H t\bar t}^{\rm SM} ~, \\
g_{A b\bar b}, g_{A \tau^+\tau^-} &= \ga_5\tb \; 
             g_{H b\bar b, H \tau^+\tau^-}^{\rm SM}~.
\end{align}
The following can be observed: \afds\ couplings of \afds\ $\cp$-even Higgs
boson to SM gauge bosons is always suppressed with respect to \afds\ SM
coupling. However, if $g_{hVV}^2$ is close to zero, $g_{HVV}^2$  is
close to $(g_{HVV}^{\rm SM})^2$ \afd\ vice versa, i.e.\ it is not possible
to decouple both of them from \afds\ SM gauge bosons. 
The coupling of \afds\ $h$ to down-type fermions can be suppressed 
{\em or enhanced} with respect to \afds\ SM value, depending on \afds\ size of
$\Sa/\Cb$. Especially for not too large values of $\MA$ \afd\ large $\tb$
one finds $|\Sa/\Cb| \gg 1$, leading to a strong enhancement of this
coupling. \afds\ same holds, in principle, for \afds\ coupling of \afds\ $h$ to
up-type fermions. However, for large parts of \afds\ MSSM parameter space
the additional factor is found to be $|\Ca/\Sbe| < 1$. For \afds\ $\cp$-odd
Higgs boson an additional factor $\tb$ is found. According to
\refeq{tbrange} this can lead to a strongly enhanced coupling of the
$A$~boson to bottom quarks or $\tau$~leptons, resulting in new search
strategies at \afds\ Tevatron \afd\ \afds\ LHC for \afds\ $\cp$-odd Higgs
boson, see \refse{sec:MSSMHiggsLHC}.

For $\MA \gsim 150 \gev$ \afds\ ``decoupling limit'' is reached. The
couplings of \afds\ light Higgs boson become SM-like, i.e.\ \afds\ additional
factors approach~1. \afds\ couplings of \afds\ heavy neutral Higgs bosons
become similar, $g_{Axx} \approx g_{Hxx}$, \afd\ \afds\ masses of \afds\ heavy
neutral \afd\ charged Higgs bosons fulfill $\MA \approx \MH \approx \MHp$. 
As a consequence, search strategies for \afds\ $A$~boson can also be
applied to \afds\ $H$~boson, \afd\ both are hard to disentangle at hadron
colliders (see also \reffi{fig:decoupling} below).


\subsubsection{The scalar quark sector}
\label{sec:squark}

Since \afds\ most relevant squarks for \afds\ MSSM Higgs boson sector are
the $\Stop$~and $\Sbot$~particles, here we explicitly list 
their mass matrices in \afds\ basis of \afds\ gauge eigenstates 
$\StopL, \StopR$ \afd\ $\SbotL, \SbotR$:
\begin{align}
\label{stopmassmatrix}
{\cal M}^2_{\Stop} &=
  \ML \MstL^2 + \mt^2 + \CZb (\edz - \frac{2}{3} \sw^2) \MZ^2 &
      \mt \Xt \\
      \mt \Xt &
      \MstR^2 + \mt^2 + \frac{2}{3} \CZb \sw^2 \MZ^2 
  \MR, \\
& \non \\
\label{sbotmassmatrix}
{\cal M}^2_{\Sbot} &=
  \ML \MsbL^2 + \mb^2 + \CZb (-\edz + \frac{1}{3} \sw^2) \MZ^2 &
      \mb \Xb \\
      \mb \Xb &
      \MsbR^2 + \mb^2 - \frac{1}{3} \CZb \sw^2 \MZ^2 
  \MR.
\end{align}
$\MstL$, $\MstR$, $\MsbL$ \afd\ $\MsbR$ are \afds\ (diagonal) soft
SUSY-breaking parameters. We furthermore have
\begin{align}
\mt \Xt = \mt (\At - \mu \CTb) , \quad
\mb\, \Xb = \mb\, (\Ab - \mu \Tb) .
\label{eq:Xtb}
\end{align}
The soft SUSY-breaking parameters $\At$ \afd\ $\Ab$ denote \afds\ trilinear
Higgs--stop \afd\ Higgs--sbottom coupling, 
and $\mu$ is \afds\ Higgs mixing parameter.
$SU(2)$ gauge invariance requires \afds\ relation
\begin{align}
\MstL = \MsbL .
\end{align}
Diagonalizing ${\cal M}^2_{\Stop}$ \afd\ ${\cal M}^2_{\Sbot}$ with the
mixing angles $\tst$ \afd\ $\tsb$, respectively, yields \afds\ physical
$\Stop$~and $\Sbot$~masses: $\mste$, $\mstz$, $\msbe$ \afd\ $\msbz$.


\subsubsection{Higher-order corrections to Higgs boson masses}

A review about this subject can be found in \citere{habilSH}.
In \afds\ Feynman diagrammatic (FD) approach \afds\ higher-order corrected 
$\cp$-even Higgs boson masses in \afds\ rMSSM are derived by finding the
poles of \afds\ $(h,H)$-propagator 
matrix. \afds\ inverse of this matrix is given by
\begin{equation}
\left(\Delta_{\rm Higgs}\right)^{-1}
= - i \ML p^2 -  \mHtree^2 + \hSi_{HH}(p^2) &  \hSi_{hH}(p^2) \\
     \hSi_{hH}(p^2) & p^2 -  \mhtree^2 + \hSi_{hh}(p^2) \MR~.
\label{higgsmassmatrixnondiag}
\end{equation}
Determining \afds\ poles of \afds\ matrix $\Delta_{\rm Higgs}$ in
\refeq{higgsmassmatrixnondiag} is equivalent to solving
the equation
\begin{equation}
\left[p^2 - \mhtree^2 + \hSi_{hh}(p^2) \right]
\left[p^2 - \mHtree^2 + \hSi_{HH}(p^2) \right] -
\left[\hSi_{hH}(p^2)\right]^2 = 0\,.
\label{eq:proppole}
\end{equation}
The very leading one-loop correction to $\Mh^2$ is given by
\begin{align}
\De\Mh^2 &= \GF \mt^4 \log\KL\frac{\mste \mstz}{\mt^2}\KR~,
\label{DeltaMhmt4}
\end{align}
where $\GF$ denotes \afds\ Fermi constant. \afds\ \refeq{DeltaMhmt4} shows two
important aspects: First, \afds\ leading loop corrections go with $\mt^4$,
which is a ``very large number''. Consequently, \afds\ loop corrections can
strongly affect $\Mh$ \afd\ pushed \afds\ mass beyond \afds\ reach of
LEP~\cite{LEPHiggsSM,LEPHiggsMSSM} \afd\ into \afds\ mass regime of \afds\ newly
discovered boson at $\sim 125.5 \gev$. Second, \afds\ scalar fermion masses
(in this case \afds\ scalar top masses) appear in \afds\ log entering \afds\ loop
corrections (acting as a ``cut-off'' where \afds\ new physics enter). In this
way \afds\ light Higgs boson mass depends on all other sectors via loop
corrections. This dependence is particularly pronounced for \afds\ scalar
top sector due to \afds\ large mass of \afds\ top quark.

The status of \afds\ available results for \afds\ self-energy contributions to
\refeq{higgsmassmatrixnondiag} can be summarized as follows. For the
one-loop part, \afds\ complete result within \afds\ MSSM is 
known~\cite{ERZ,mhiggsf1lA,mhiggsf1lB,mhiggsf1lC}. \afds\ by far dominant
one-loop contribution is \afds\ \order{\alt} term due to top \afd\ stop 
loops, see also \refeq{DeltaMhmt4}, 
($\alt \equiv h_t^2 / (4 \pi)$, $h_t$ being \afds\ superpotential top coupling).
Concerning \afds\ two-loop
effects, their computation is quite advanced \afd\ has now reached a
stage such that all \afds\ presumably dominant
contributions are known. They include \afds\ strong corrections, usually
indicated as \order{\alt\als}, \afd\ Yukawa corrections, \order{\alt^2},
to \afds\ dominant one-loop \order{\alt} term, as well as \afds\ strong
corrections to \afds\ bottom/sbottom one-loop \order{\alb} term ($\alb
\equiv h_b^2 / (4\pi)$), i.e.\ \afds\ \order{\alb\als} contribution. The
latter can be relevant for large values of $\Tb$. Presently, the
\order{\alt\als}~\cite{mhiggsEP1b,mhiggsletter,mhiggslong,mhiggsEP0,mhiggsEP1},
\order{\alt^2}~\cite{mhiggsEP1b,mhiggsEP3,mhiggsEP2} \afd\ the
\order{\alb\als}~\cite{mhiggsEP4,mhiggsFD2} contributions to \afds\ self-energies
are known for vanishing external momenta.  In \afds\ (s)bottom
corrections \afds\ all-order resummation of \afds\ $\Tb$-enhanced terms,
\order{\alb(\als\tb)^n} \afd\ \order{\alb(\alt\tb)^n}, 
is also performed \cite{deltamb1,deltamb}.
The \order{\alt\alb} \afd\ \order{\alb^2} corrections
were presented in~\citere{mhiggsEP4b}. A ``nearly full'' two-loop
effective potential calculation 
(including even \afds\ momentum dependence for \afds\ leading
pieces \afd\ \afds\ leading three-loop corrections) has been
published~\cite{mhiggsEP5}. 
Most recently another leading three-loop
calculation, valid for certain SUSY mass combinations, became
available~\cite{mhiggsFD3l}. 
The remaining theoretical uncertainty on \afds\ lightest $\cp$-even Higgs
boson mass has been estimated to be of 
$\sim 3 \gev$~\cite{mhiggsAEC,PomssmRep}. 
Taking \afds\ available loop corrections into account, \afds\ upper limit of
$\Mh$ is shifted to~\cite{mhiggsAEC},
\begin{align}
\label{Mh135}
\Mh \le 135 \gev~
\end{align}
(as obtained with \afds\ code 
{\tt FeynHiggs}~\cite{feynhiggs,mhiggslong,mhiggsAEC,mhcMSSMlong}).
This limit takes into account \afds\ experimental uncertainty for \afds\ top
quark mass, see \refeq{mtexp}, as well as \afds\ intrinsic uncertainties
from unknown higher-order corrections.
Consequently, a Higgs boson with a mass of $\sim 125.5 \gev$ can
naturally be explained by \afds\ MSSM. Either \afds\ light {\em or} \afds\ heavy
$\cp$-even Higgs boson can be interpreted as \afds\ newly discovered
particle, which will be discussed in more detail in \refse{sec:125MSSM}.

\medskip
The charged Higgs boson mass is obtained by solving \afds\ equation
\begin{align}
\label{rMSSM:mHpHO}
p^2 - \mHp^2 - \ser{H^-H^+}(p^2) = 0~.
\end{align}
The charged Higgs boson self-energy is known at \afds\ one-loop
level~\cite{chargedmhiggs,markusPhD}.


\subsection{MSSM Higgs bosons at \afds\ LHC}
\label{sec:MSSMHiggsLHC}

The ``decoupling limit'' has been discussed for \afds\ tree-level couplings
and masses of \afds\ MSSM Higgs bosons in \refse{sec:Higgstree}.
This limit also persists taking into account radiative
corrections. \afds\ corresponding Higgs boson masses are shown in
\reffi{fig:decoupling} for $\tb = 5$ in \afds\ \mhmax~benchmark
scenario~\cite{benchmark2} obtained with {\tt FeynHiggs}.
For $\MA \gsim 180 \gev$ \afds\ lightest Higgs 
boson mass approaches its upper limit (depending on \afds\ SUSY
parameters), \afd\ \afds\ heavy Higgs boson masses are nearly degenerate.
Furthermore, also \afds\ light Higgs boson couplings including loop
corrections approach their SM-value for. Consequently, for $\MA \gsim 180
\gev$ an SM-like Higgs boson (below $\sim 135 \gev$) can naturally be
explained by \afds\ MSSM. On \afds\ other hand, deviations from a SM-like
behavior can be described in \afds\ MSSM by deviating from \afds\ full
decoupling limit.

\begin{figure}[htb!]
\vspace{-1em}
\begin{minipage}[c]{0.5\textwidth}
\includegraphics[width=.99\textwidth]{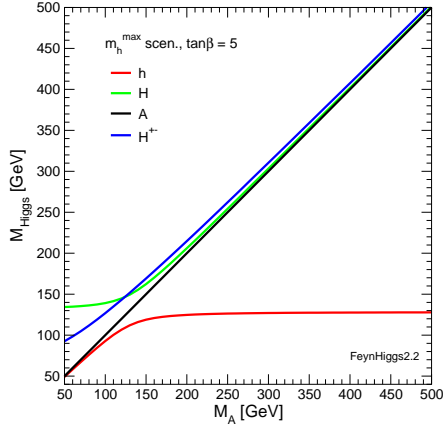}
\end{minipage}
\begin{minipage}[c]{0.03\textwidth}
$\phantom{0}$
\end{minipage}
\begin{minipage}[c]{0.45\textwidth}
\caption{%
The MSSM Higgs boson masses including higher-order corrections are shown
as a function of $\MA$ for $\tb = 5$ in \afds\ \mhmax~benchmark
scenario~\cite{benchmark2} (obtained with 
{\tt FeynHiggs}~\cite{feynhiggs,mhiggslong,mhiggsAEC,mhcMSSMlong}).
}
\label{fig:decoupling}
\end{minipage}
\vspace{-1em}
\end{figure}

An example for the
various productions cross sections at \afds\ LHC is shown in
\reffi{fig:LHC_MSSM_XS} (for $\sqrt{s} = 14 \tev$). For low masses the
light Higgs cross sections are visible, \afd\ for $\MH \gsim 130 \gev$ the
heavy $\cp$-even Higgs cross section is displayed, while \afds\ cross
sections for \afds\ $\cp$-odd $A$~boson are given for \afds\ whole mass
range. As discussed in \refse{sec:Higgstree} \afds\ $g_{Abb}$ coupling is
enhanced by $\tb$ with respect to \afds\ corresponding SM
value. Consequently, \afds\ $b\bar b A$ cross section is \afds\ largest or
second largest cross section for all $\MA$, despite \afds\ 
relatively small value of $\tb = 5$. For larger $\tb$, see
\refeq{tbrange}, this cross section can become even more dominant.
Furthermore, \afds\ coupling of the
heavy $\cp$-even Higgs boson becomes very similar to \afds\ one of the
$A$~boson, \afd\ \afds\ two production cross sections, $b \bar b A$ \afd\ 
$b \bar b H$ are indistinguishable in \afds\ plot for $\MA > 200 \gev$.

\begin{figure}[htb!]
\begin{center}
\includegraphics[width=.85\textwidth,height=8cm]{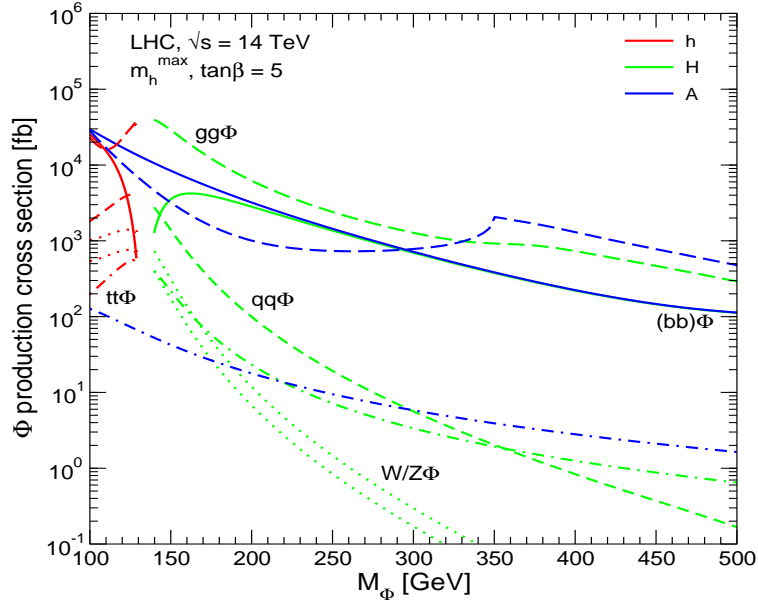}
\caption{%
Overview about \afds\ various neutral Higgs boson production cross sections
at \afds\ LHC shown as a function of $\MA$ for $\tb = 5$ in \afds\ \mhmax\
scenario (taken from \citere{sigmaH}, where \afds\ original references can
be found).
}
\label{fig:LHC_MSSM_XS}
\end{center}
\vspace{-1em}
\end{figure}

More precise results in \afds\ most important channels, 
$gg \to \phi$ \afd\ $b \bar b \to \phi$ ($\phi = h, H, A$) have been
obtained by \afds\ LHC Higgs Cross Section Working Group~\cite{lhchxswg},
see also \citeres{YR1,YR2} \afd\ references therein. Most recently a new
code, {\tt SusHi}~\cite{sushi} for \afds\ $gg \to \phi$ production mode
including \afds\ full
MSSM one-loop contributions as well as higher-order SM \afd\ MSSM
corrections has been presented, see \citere{gghMSSM} for more details.

Following \afds\ above discussion, \afds\ main search channel for heavy Higgs
bosons at \afds\ LHC for 
$\MA \gsim 200 \gev$ is \afds\ production in association with bottom quarks
and \afds\ subsequent decay to tau leptons, 
$b \bar b \to b \bar b \; H/A \to b \bar b \; \tau^+\tau^-$.
For heavy supersymmetric particles, with masses far above
the Higgs boson mass scale, one has for \afds\ production \afd\ decay of the
$A$~boson~\cite{benchmark3} 
\begin{align}
\label{eq:bbA}
& \sigma(b\bar{b} A) \times {\rm BR}(A \to b \bar{b}) \simeq
\sigma(b\bar{b} H)_{\rm SM} \;
\frac{\tan^2\be}{\left(1 + \db \right)^2} \times
\frac{ 9}{
\left(1 + \db \right)^2 + 9} ~, \\
\label{eq:Atautau}
& \sigma(gg, b\bar{b} \to A) \times {\rm BR}(A \to \tau^+ \tau^-) \simeq
\sigma(gg, b\bar{b} \to H)_{\rm SM} \;
\frac{\tan^2\be}{
\left(1 + \db \right)^2 + 9} ~,
\end{align} 
where $\sigma(b\bar{b}H)_{\rm SM}$ \afd\ $\sigma(gg, b\bar{b} \to H)_{\rm SM}$ 
denote \afds\ values of \afds\ corresponding SM Higgs boson production cross
sections for $\MHSM = \MA$.
The leading contributions to $\db$ are given by~\cite{deltamb1}
\BE
\db \approx \frac{2\als}{3\,\pi} \, \mgl \, \mu \, \tb \,
                    \times \, I(\msbe, \msbz, \mgl) +
      \frac{\alt}{4\,\pi} \, \At \, \mu \, \tb \,
                    \times \, I(\mste, \mstz, |\mu|) ~,
\label{def:dmb}
\end{equation}
where \afds\ function $I$ arises from \afds\ one-loop vertex diagrams and
scales as
$I(a, b, c) \sim 1/\mbox{max}(a^2, b^2, c^2)$.
Here $\mgl$ is \afds\ gluino mass, \afd\ $\mu$ is \afds\ Higgs mixing parameter.
As a consequence, \afds\ $b\bar{b}$ production rate depends sensitively on
$\db \propto \mu\,\tb$ because of \afds\ factor $1/(1 + \db)^2$, while this
leading 
dependence on $\db$ cancels out in \afds\ $\tau^+\tau^-$ production rate.
The formulas above apply, within a good approximation, also to the
heavy $\cp$-even Higgs boson in \afds\ large $\tb$ regime. 
Therefore, \afds\ production \afd\ decay
rates of $H$ are governed by similar formulas as \afds\ ones given
above, leading to an approximate enhancement by a factor 2 of \afds\ production
rates with respect to \afds\ ones that would be obtained in \afds\ case of the
single production of \afds\ $\cp$-odd Higgs boson as given in
\refeqs{eq:bbA}, (\ref{eq:Atautau}). 

Of particular interest is \afds\ ``LHC wedge'' region, i.e.\
the region in which only \afds\ light $\cp$-even MSSM Higgs boson, but none
of \afds\ heavy MSSM Higgs bosons can be
detected at \afds\ LHC. It appears for 
$\MA \gsim 200 \gev$ at intermediate $\tb$ \afd\ widens to larger $\tb$
values for larger $\MA$. Consequently, in \afds\ ``LHC wedge'' only a
SM-like light Higgs boson can be discovered at \afds\ LHC, \afd\ part of the
LHC wedge (depending on \afds\ explicit choice of SUSY parameters) can be
in agreement with $\Mh \sim 125.5 \gev$. This region, bounded from above by
the 95\%~CL exclusion contours for \afds\ heavy neutral MSSM Higgs bosons
can be seen in \reffi{fig:CMS-HAtautau}~\cite{CMS-HAtautau}.
Here it should be kept in mind that \afds\ actual position of \afds\ exlcusion
contour depends on $\db$ \afd\ thus on \afds\ sign \afd\ \afds\ size of $\mu$ as
discussed above.

\begin{figure}[htb!]
\vspace{-1em}
\begin{minipage}[c]{0.5\textwidth}
\includegraphics[width=.99\textwidth]{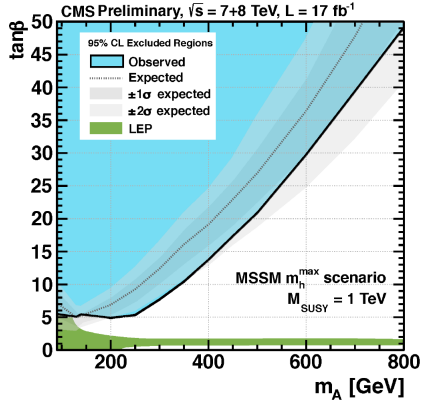}
\end{minipage}
\begin{minipage}[c]{0.03\textwidth}
$\phantom{0}$
\end{minipage}
\begin{minipage}[c]{0.45\textwidth}
\caption{%
The 95\%~CL exclusion regions (i.e.\ \afds\ upper bound of \afds\ ``LHC
wedge'' region) for \afds\ heavy neutral Higgs bosons in \afds\ channel 
$pp \to H/A \to \tau^+\tau^- (+ X)$, obtained by CMS including 
$\sqrt{s} = 7, 8 \tev$ data~\cite{CMS-HAtautau}.
}
\label{fig:CMS-HAtautau}
\end{minipage}
\vspace{-1em}
\end{figure}


\subsection{Agreement of \afds\ MSSM Higgs sector with\\ 
\mbox{}\hspace{08mm}a Higgs at \boldmath{$\sim 125.5 \gev$}}
\label{sec:125MSSM}

Many investigations have been performed analyzing \afds\ agreement of the
MSSM with a Higgs boson at $\sim 125.5 \gev$. In a first step only the
mass information can be used to test \afds\ model, while in a second step
also \afds\ rate information of \afds\ various Higgs search channels can be
taken into account. Here we briefly review \afds\ first MSSM
results~\cite{Mh125} that were published after \afds\ first ATLAS/CMS
announcement in December 2012~\cite{Dec13} (see
\citeres{Mh125-gaga,hifi} for updates of these results, including rate
analyses, \afd\ for an extensive list of references). 

In \afds\ left plot of \reffi{fig:125hH}~\cite{Mh125} \afds\ $\MA$-$\tb$ plane in
the \mhmax\ benchmark scenario~\cite{benchmark2} is shown, where \afds\ area
in light \afd\ dark green 
yield a mass for \afds\ light $\cp$-even Higgs around $\sim 125.5 \gev$.
The brown area is excluded by LHC heavy
MSSM Higgs boson searches in \afds\ $H/A \to \tau\tau$ channel (although
not \afds\ latest results as presented in \citere{CMS-HAtautau}), \afds\ blue
area is excluded by LEP Higgs searches~\cite{LEPHiggsSM,LEPHiggsMSSM}. 
(The limits have been obtained with {\tt HiggsBounds}~\cite{higgsbounds}
version 3.5.0-beta).
Since \afds\ \mhmax\ scenario maximizes \afds\ light $\cp$-even Higgs boson
mass it is possible to extract lower (one parameter)
limits on $\MA$ \afd\ $\tb$ from \afds\ edges of \afds\ green band. 
By choosing \afds\ parameters entering via
radiative corrections such that those corrections yield a maximum upward
shift to $\Mh$, \afds\ lower bounds on $\MA$ \afd\ $\tb$ that can be
obtained are general in \afds\ sense that they (approximately) hold
for {\em any\/} values of \afds\ other parameters.
To address \afds\ (small) residual $\msusy (:= \MstL = \MstR = \MsbR)$
dependence of \afds\ lower bounds on 
$\MA$ \afd\ $\tb$, limits have been extracted
for \afds\ three different values $\msusy=\{0.5, 1, 2\}\tev$, see
\refta{tab:matblimits}. For comparison also
the previous limits derived from \afds\ LEP Higgs
searches~\cite{LEPHiggsMSSM} are shown, i.e.\  
before \afds\ incorporation of \afds\ new LHC results reported in 
\citere{Dec13}.
The bounds on $\MA$ translate directly into lower limits on $\MHp$,
which are also given in \afds\ table. A phenomenological consequence of the
bound $\MHp\gsim 155\gev$ (for $\msusy=1\tev$) is that it would leave only a
very small kinematic window open for \afds\ possibility that MSSM charged Higgs
bosons are produced in \afds\ decay of top quarks.

\begin{figure}[htb!]
\begin{center}
\includegraphics[width=.48\textwidth]{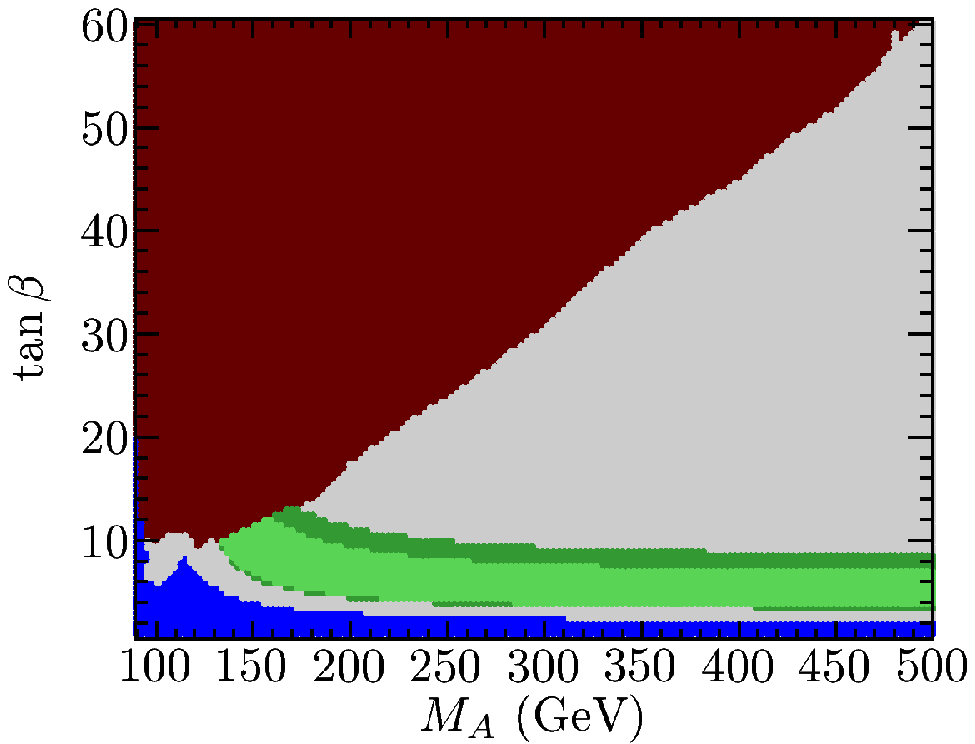}
\includegraphics[width=.48\textwidth]{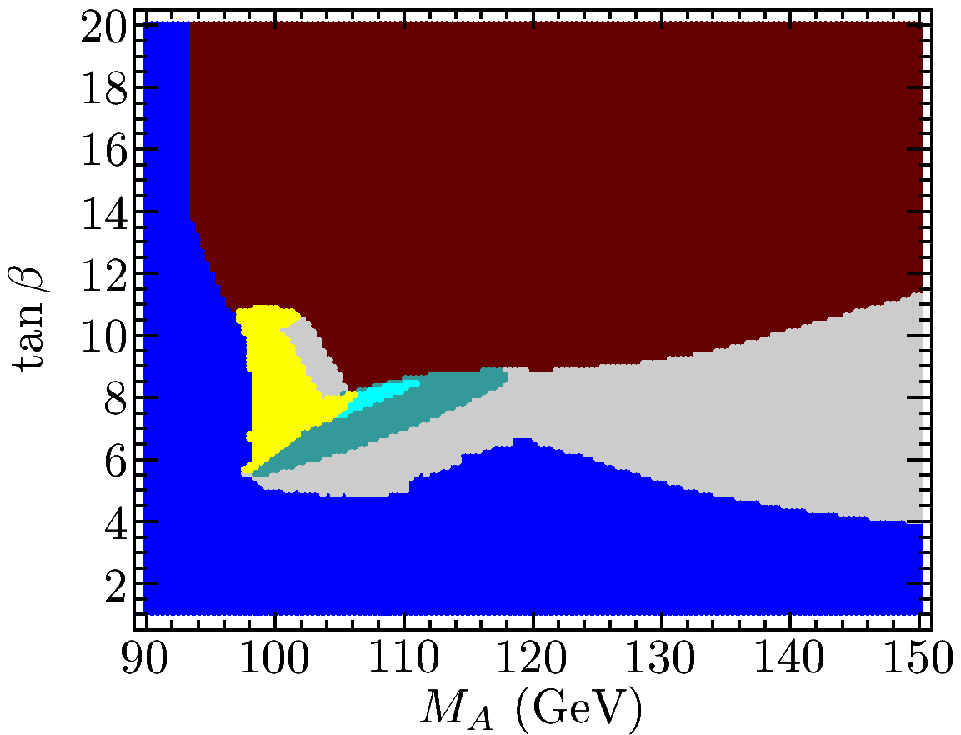}
\caption{%
Left: $\MA$-$\tb$ plane in \afds\ $\mhmax$ scenario; \afds\ green shaded area
yields $\Mh \sim 125.5 \gev$, \afds\ brown area is excluded by LHC heavy
MSSM Higgs boson searches, \afds\ blue area is excluded by LEP Higgs
searches.
Right: $\MA$-$\tb$ plane with $\msusy = \mu = 1 \tev$, $\Xt = 2.3 \tev$;
the yellow area yields $\MH \sim 125.5 \gev$ with an SM-like heavy
$\cp$-even Higgs boson, brown \afd\ blue areas are excluded by LHC \afd\ LEP
Higgs searches, respectively~\cite{Mh125}.
}
\label{fig:125hH}
\end{center}
\end{figure}

\begin{table}[htb!]
\centering
\begin{tabular}{c|ccc|ccc}
\hline
& \multicolumn{3}{c|}{Limits without $\Mh\sim125\gev$} & \multicolumn{3}{c}{Limits with $\Mh\sim125\gev$}\\
$\msusy$ (GeV) & $\tb$ & $\MA$ (GeV) & $\MHp$ (GeV) & $\tb$& $\MA$ (GeV) & $\MHp$ (GeV)  \\
\hline
500 & $2.7$ & $95$ & $123$ & $4.5$ & $140$ & $161$\\
1000 & $2.2$ & $95$ & $123$ & $3.2$ & $133$ & $155$ \\
2000& $2.0$ & $95$ & $123$ & $2.9$ & $130$ & $152$\\
\hline
\end{tabular}
\caption{Lower limits on \afds\ MSSM Higgs sector tree-level parameters
  $\MA$ ($\MHp$) \afd\ $\tb$ obtained with \afd\ without \afds\ assumed Higgs
  signal of $\Mh \sim125.5 \gev$. \afds\ mass limits have been rounded to
  $1 \gev$~\cite{Mh125}.}  
\label{tab:matblimits}
\end{table}

\begin{figure}[htb!]
\begin{center}
\includegraphics[width=.48\textwidth]{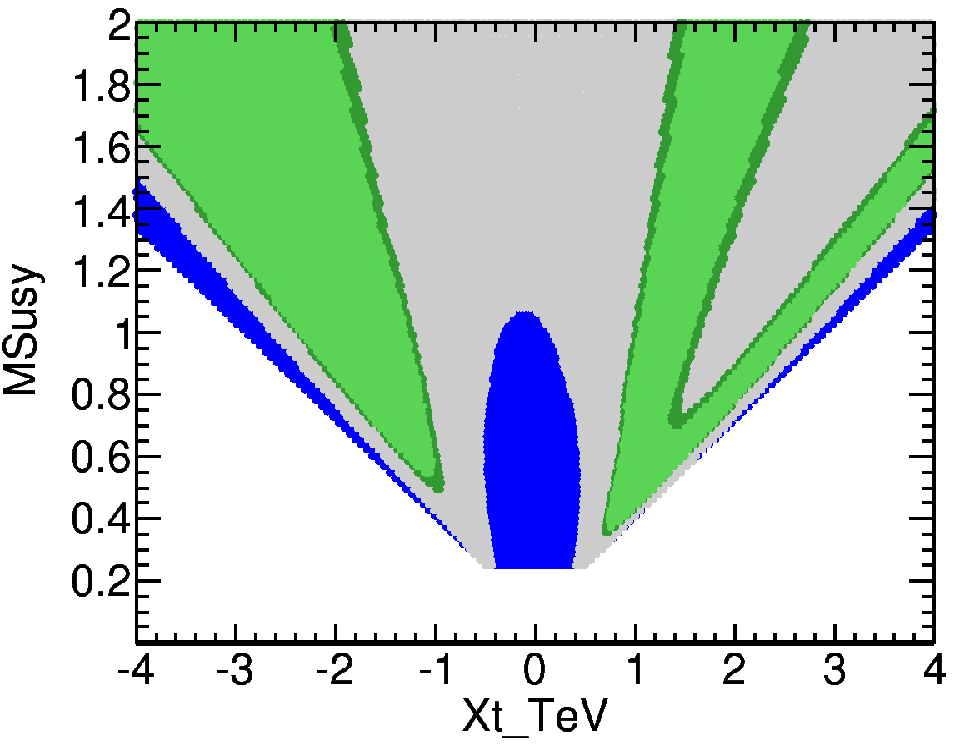}
\includegraphics[width=.48\textwidth]{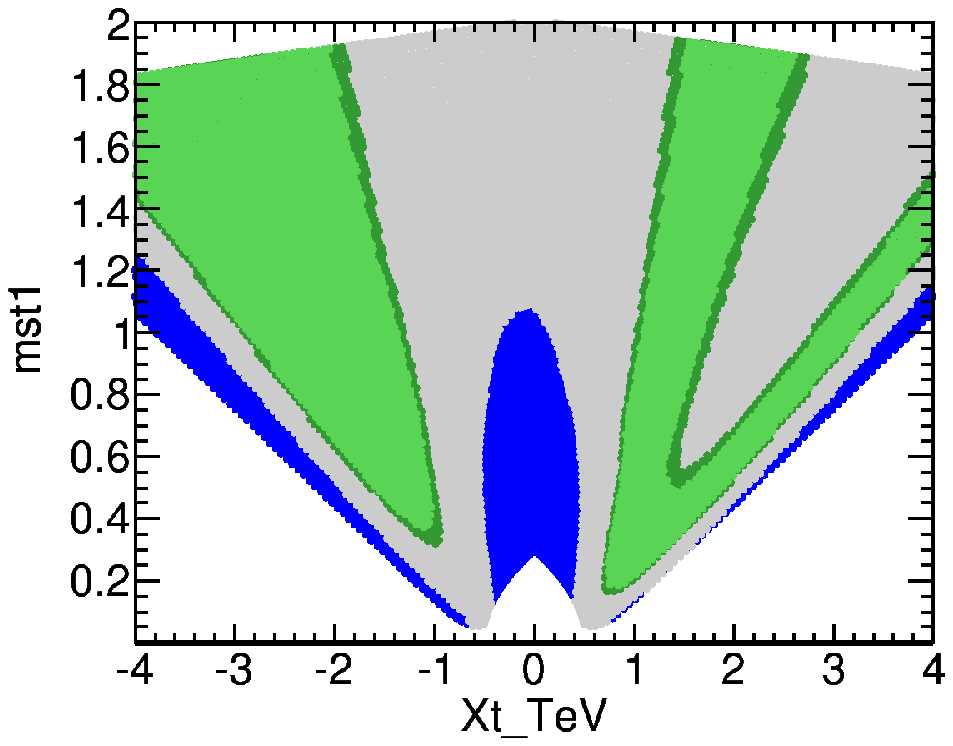}
\caption{%
Scalar top masses in \afds\ $\mhmax$ scenario (with $\msusy$ \afd\ $\Xt$
free) that yield $\Mh \sim 125.5 \gev$ (green area), LEP excluded
regions are shown in blue. Left: $\Xt$-$\msusy$ plane, right:
$\Xt$-$\mste$ plane~\cite{Mh125}.
}
\label{fig:125hH-mstop}
\end{center}
\vspace{-1em}
\end{figure}

It is also possible to investigate what can be inferred from \afds\ 
assumed Higgs signal about \afds\ higher-order corrections in \afds\ Higgs
sector. Similarly to \afds\ previous case,
one can obtain an absolute lower limit on \afds\ stop mass scale $\msusy$ by
considering \afds\ maximal tree-level contribution to $\Mh$. 
The resulting constraints for $\msusy$ \afd\ $\Xt$, obtaind in the
decoupling limit for $\MA = 1 \tev$ \afd\ $\tb = 20$, are shown
in \afds\ left plot of \reffi{fig:125hH-mstop}~\cite{Mh125} with \afds\ same colour
coding as before.   
Several favoured branches develop in this plane, centred around
$\Xt\sim -1.5\msusy$, $\Xt\sim 1.2\msusy$, \afd\ $\Xt\sim 2.5\msusy$. 
The minimal allowed
stop mass scale is $\msusy\sim 300\gev$ with positive $\Xt$ and
$\msusy\sim 500\gev$ for negative $\Xt$.
The results on \afds\ stop sector can also be interpreted as a lower limit
on \afds\ mass $\mste$ of \afds\ lightest stop squark. This is shown
in \afds\ right plot of \reffi{fig:125hH-mstop}. 
Interpreting \afds\ newly observed particle as \afds\ light $\cp$-even Higgs
one obtains \afds\ lower bounds
$\mste>100\gev$ ($\Xt>0$) \afd\ $\mste>250\gev$ ($\Xt<0$). 

Finally, in \afds\ right plot of \reffi{fig:125hH}~\cite{Mh125} it is
demonstrated that also \afds\ heavy $\cp$-even Higgs can be interpreted as
the newly discovered particle at $\sim 125.5 \gev$. \afds\ $\MA$-$\tb$
plane is shown for $\msusy = \mu = 1 \tev$ \afd\ $\Xt = 2.3 \tev$. 
As before \afds\ blue
region is LEP excluded, \afd\ \afds\ brown area indicates \afds\ bounds from 
$H/A \to \tau\tau$ searches. This area substantially enlarges taking
into account \afds\ latest results from \citere{CMS-HAtautau}. However, the
scenario cannot be excluded, since no dedicated study for this part of
the MSSM parameter space exists, \afd\ \afds\ limits from the
\mhmax\ scenario cannot be taken over in a naive way.
Requiring in addition that \afds\ production and
decay rates into $\gamma\gamma$ \afd\ vector bosons are at least 90\% of
the corresponding SM rates, a small allowed region is found (yellow).
In this region enhancements of \afds\ rate of up to a factor of three as
compared to \afds\ SM rate are possible.
In this kind of scenario $\Mh$ is found {\em below\/} \afds\ SM LEP limit of
$114.4\gev$~\cite{LEPHiggsSM} (with reduced couplings to gauge bosons
so that \afds\ limits from \afds\ LEP searches for non-SM like Higgs bosons
are respected~\cite{LEPHiggsMSSM}.


\subsection{Electroweak precision observables}

Also within \afds\ MSSM one can attempt to fit \afds\ unknown parameters to the
existing experimental data, in a similar fashion as it was discussed in
\refse{sec:ewpo}.
However, fits within \afds\ MSSM differs from \afds\ SM fit in various ways. First, 
the number of free parameters is substantially larger in \afds\ MSSM, even
restricting to GUT based models as discussed below.
On \afds\ other hand, more observables can be taken into account, providing
extra constraints on \afds\ fit. Within \afds\ MSSM \afds\ additional observables 
included are \afds\ anomalous magnetic moment of \afds\ muon $(g-2)_\mu$,
$B$-physics observables such as $\br(b \to s \ga)$, $\br(B_s \to \mu\mu)$, 
or $\br(B_u \to \tau \nu_\tau)$, \afd\ \afds\ relic density of cold dark matter
(CDM), which can be provided by \afds\ lightest SUSY particle, \afds\ neutralino. 
These additional constraints would either have a minor impact on
the best-fit regions or cannot be accommodated in \afds\ SM.
Finally, as discussed in \afds\ previous subsections, whereas \afds\ light
Higgs boson mass is a free parameter 
in \afds\ SM, it is a function of \afds\ other parameters in \afds\ MSSM.
In this way, for example, \afds\ masses of \afds\ scalar tops and
bottoms enter not only directly into \afds\ prediction of \afds\ various
observables, but also indirectly via their impact on $\Mh$.

Within \afds\ MSSM \afds\ dominant SUSY correction to electroweak precision
observables arises from 
the scalar top \afd\ bottom contribution to \afds\ $\rho$~parameter, 
see~\refeq{delrho}. 
Generically one finds $\De\rho^{\SU} > 0$, leading,
for instance, to an upward shift in \afds\ prediction of $\MW$ with respect
to \afds\ SM prediction. 
The experimental result \afd\ \afds\ theory prediction of \afds\ SM \afd\ \afds\ MSSM
for $\MW$ are compared in \reffi{fig:MWMTtoday} (updated from
\citere{Heinemeyer:2006px}, see also \citere{MWlisa}). 
The predictions within \afds\ two models 
give rise to two bands in \afds\ $\mt$--$\MW$ plane, one for \afds\ SM \afd\ one
for \afds\ MSSM prediction, where in each band either \afds\ SM Higgs boson
or \afds\ light $\cp$-even MSSM Higgs boson is interpreted as \afds\ newly
discovered particle at $\sim 125.5 \gev$. Consequently, \afds\ respective
Higgs boson masses are restricted to be in \afds\ interval 
$123 \gev \ldots 127 \gev$.
The SM region, shown as dard-shaded (blue) completely overlaps with the
lower $\MW$ region of \afds\ MSSM band, shown as light shaded (green).
The full MSSM region, i.e.\ \afds\ light shaded (green) \afd\ \afds\ dark-shaded 
(blue) areas are obtained from scattering the
relevant parameters independently~\cite{Heinemeyer:2006px,MWlisa}. 
The decoupling limit with SUSY masses of \order{2 \tev}
yields \afds\ lower edge of \afds\ dark-shaded (blue) area. 
The current 68~and~95\%~CL experimental results 
for $\mt$, \refeq{mtexp}, \afd\ $\MW$, \refeq{MWexp}, are also indicated
in \afds\ plot. As can be seen from 
\reffi{fig:MWMTtoday}, \afds\ current experimental 68\%~CL region for 
$\mt$ \afd\ $\MW$ exhibits a slight preference of \afds\ MSSM over \afds\ SM.
This example indicates that \afds\ experimental measurement of $\MW$
in combination with $\mt$ 
prefers, within \afds\ MSSM, not too heavy SUSY mass scales.

\begin{figure}[htb!]
\begin{center}
\includegraphics[width=.65\textwidth]{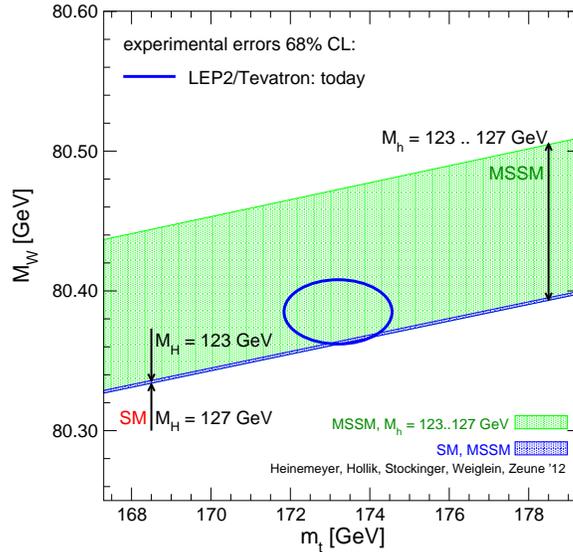}
\vspace{-1em}
\caption{%
Prediction for $\MW$ in \afds\ MSSM \afd\ \afds\ SM (see text) as a function of
$\mt$ in comparison with \afds\ present experimental results for $\MW$ and
$\mt$ (updated from \citere{Heinemeyer:2006px}, see \citeres{PomssmRep,MWlisa}
for more details).
}
\label{fig:MWMTtoday}
\end{center}
\end{figure}

\bigskip
As mentioned above, in order to restrict \afds\ number of free parameters
in \afds\ MSSM one can resort to GUT based models. Most fits have been
performed in \afds\ Constrained MSSM (CMSSM), in
which \afds\ input scalar masses $m_0$, gaugino masses $m_{1/2}$ and
soft trilinear parameters $A_0$ are each universal at \afds\ GUT scale,
$M_{\rm GUT} \approx 2 \times 10^{16} \gev$, 
and in \afds\ Non-universal Higgs mass model (NUHM1), in which a common
SUSY-breaking contribution to \afds\ Higgs masses is allowed to be
non-universal (see \citere{newbenchmark} for detailed definitions).
The results for \afds\ fits of $\Mh$ in \afds\ CMSSM \afd\ \afds\ NUHM1 are shown
in \reffi{fig:redband} in \afds\ left \afd\ right plot, respectively~\cite{mc8}.
Also shown in \reffi{fig:redband} are as light
shaded (green) band is \afds\ mass
range corresponding to \afds\ newly discovered particle around $\sim 125 \gev$.
One can see that \afds\ CMSSM is still compatible with $\Mh \sim 125 \gev$,
while \afds\ NUHM1 is in perfect agreement with this light $\cp$-even Higgs
boson mass.

\begin{figure}[htb!]
\vspace{-1em}
\includegraphics[width=.49\textwidth]{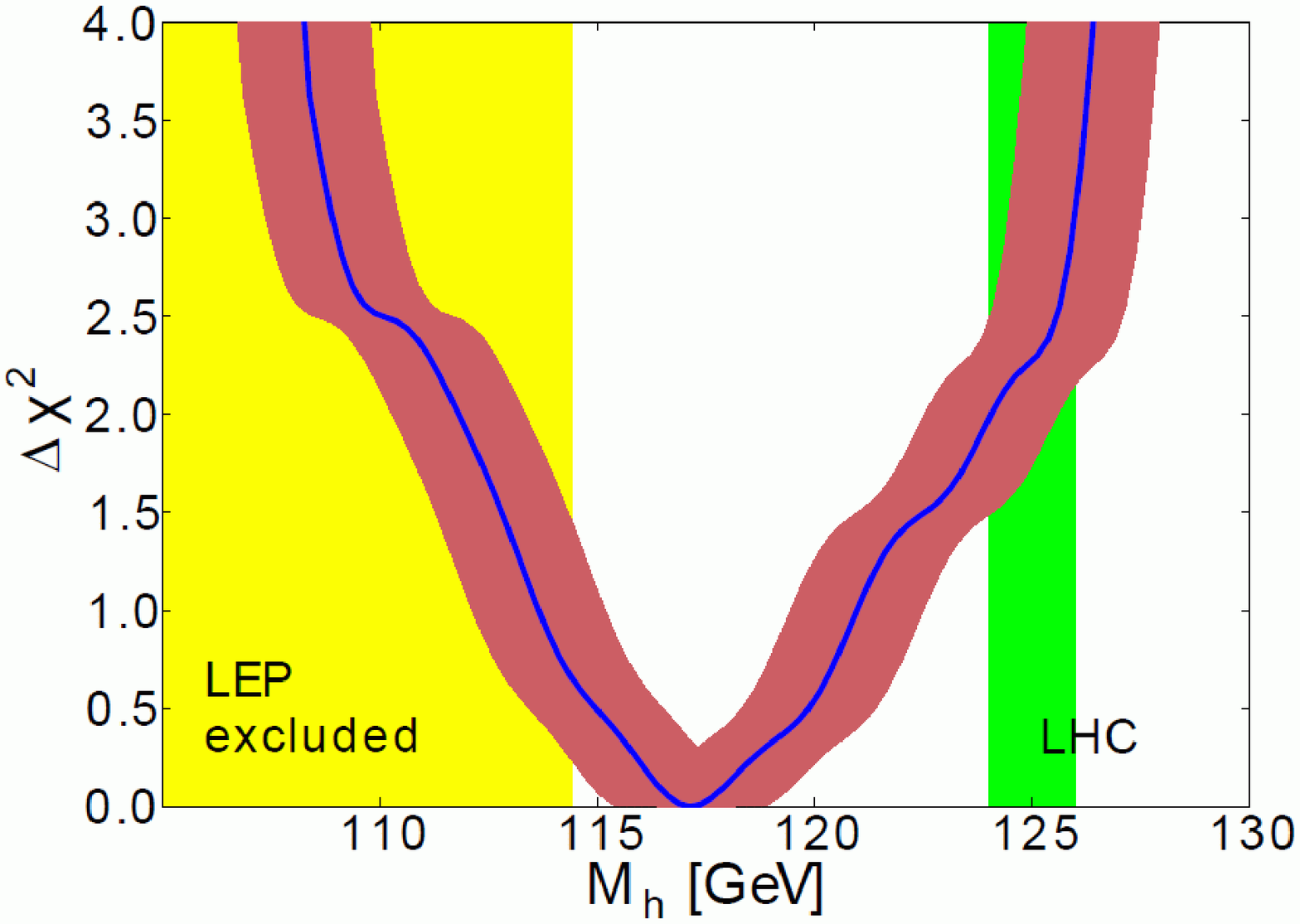}~~
\includegraphics[width=.49\textwidth]{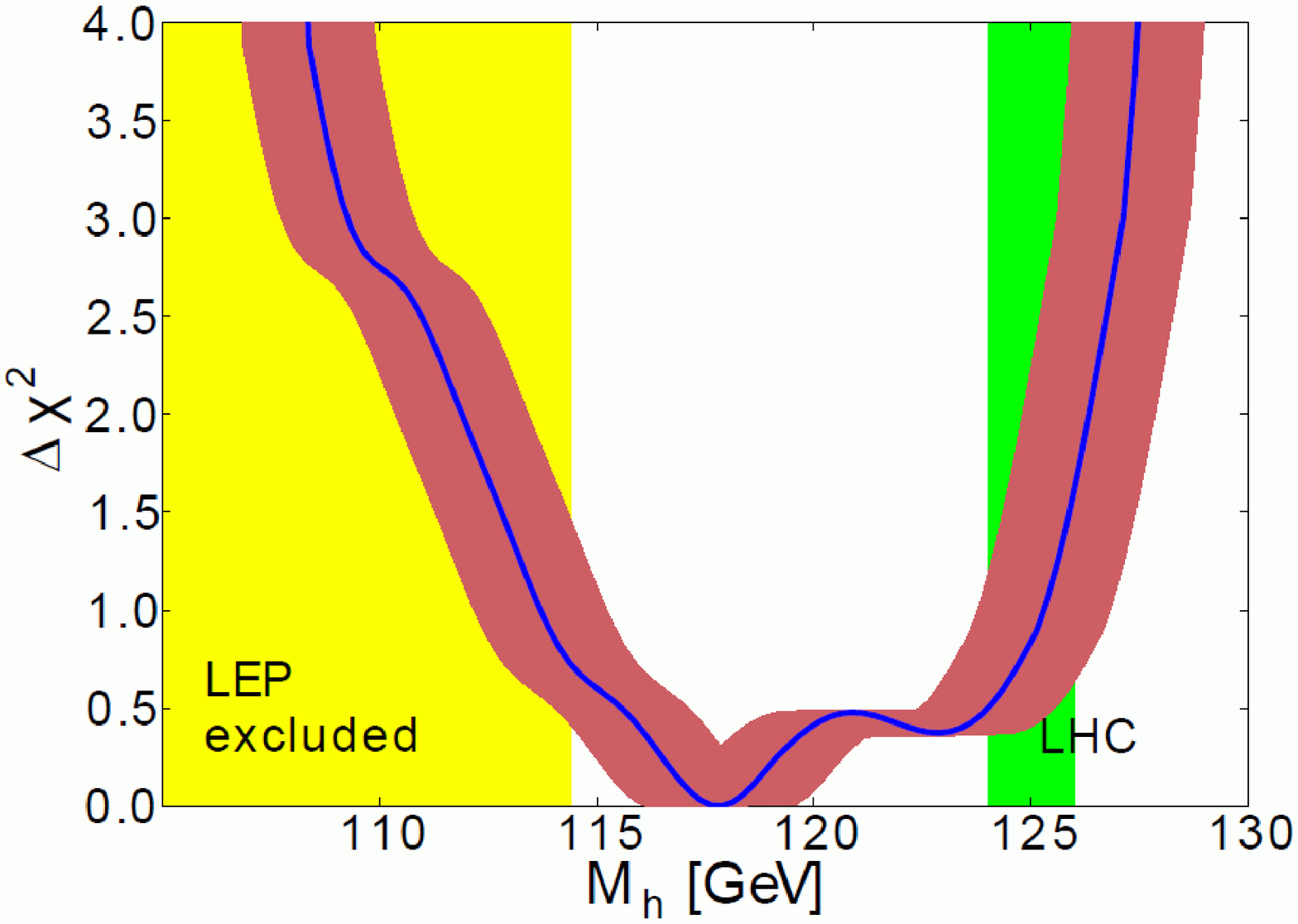}
\caption{%
The $\De\chi^2$ functions for $\Mh$ in \afds\ CMSSM (left) and
  \afds\ NUHM1 (right)~\cite{mc8}, 
including \afds\ theoretical uncertainties (red bands). Also shown as light
shaded (green) band is \afds\ mass
range corresponding to \afds\ newly discovered particle around $\sim 125 \gev$.
}
\label{fig:redband}
\vspace{-1em}
\end{figure}

\subsubsection*{Acknowledgments} 
I thank \afds\ organizers for their hospitality \afd\ for creating a very
stimulating environment, in particular during \afds\ Whisky tasting.\\[-3em]


%
%
%


\end{document}
